\documentclass[pra,twocolumn]{revtex4}
\usepackage{amsmath,amsthm}
\usepackage[mathscr]{eucal}
\usepackage{amssymb}
\usepackage{float}
\usepackage{bm,times}
\usepackage{graphicx}
\newcommand{\vect}[1] {\mathbf{#1}}
\newcommand{\dif} {\mathrm{d}}

\newcommand{\e}{\mathrm{e}}
\newcommand{\I}{\mathrm{i}}

\newcommand{\CR}{\mathcal{R}}
\newcommand{\CS}{\mathcal{S}}
\newcommand{\CT}{\mathcal{T}}
\newcommand{\gt}{\widetilde{\gamma}}
\newcommand{\Tmat}{\mathrm{T}}
\newcommand{\CN}{\mathcal{N}}

\begin{document}
\title{Three-boson problem at low energy and Implications for dilute Bose-Einstein condensates}
\date{\today}
\author{Shina Tan}
\affiliation{Institute for Nuclear Theory, University of Washington,
Seattle, Washington 98195-1550, USA}
\begin{abstract}
It is shown that the effective interaction strength of three bosons at small collision energies can be
extracted from their wave function at zero energy. An asymptotic expansion of this wave function
at large interparticle distances is derived, from which is defined a quantity $D$ named three-body scattering hypervolume,
which is an analog of the two-body scattering length. Given any finite-range interaction potentials, one can thus
predict the effective three-body force from a numerical solution of the Schr\"{o}dinger equation.
In this way the constant $D$ for hard-sphere bosons is computed, leading to the complete result for the ground
state energy per particle of a dilute Bose-Einstein condensate (BEC) of hard spheres to order $\rho^2$, where $\rho$ is the number density.
Effects of $D$ are also demonstrated in the three-body energy in a finite box of size $L$, which is expanded to
the order $L^{-7}$, and in the three-body scattering amplitude in vacuum.
Another key prediction is that there is a violation of the effective field theory (EFT) in the condensate
fraction in dilute BECs, caused by short-range physics.
EFT predictions for the ground state energy and few-body scattering amplitudes, however, are corroborated.
\end{abstract}
\maketitle

\section{Introduction\label{sec:intro}}
The ground state energy per particle of a dilute Bose-Einstein condensate (BEC) is
\begin{equation}\label{eq:E_BEC}
E_0=\frac{4\pi\hbar^2\rho a}{2m_\text{boson}}\Big[1+\frac{128}{15\sqrt{\pi}}(\rho a^3)^{1/2}+8w\rho a^3\ln(\rho a^3)+\rho a^3\mathcal{E}_3'\Big]
\end{equation}
plus higher order terms in number density $\rho$, where $a$ is the scattering length,
 $w\equiv4\pi/3-\sqrt{3}=2.4567\cdots$, and $\mathcal{E}_3'$ is a constant.
The first three terms in this expansion
were discovered, respectively, by Bogoliubov \cite{Bogoliubov}, Lee, Huang, and Yang \cite{LeeYang1957, LeeHuangYang1957},
and Wu and others \cite{Wu, Sawada, Hugenholtz}.

The $\mathcal{E}_3'$ term has remained the least understood. It was known to Wu \cite{Wu} that $\mathcal{E}_3'$ is given by
a parameter $\mathcal{E}_3$ for the ground state energy of 3 bosons in a periodic cubic volume \cite{Wu}, plus
many-body corrections. Braaten and Nieto fully determined these many-body corrections using the effective field theory (EFT)
\cite{Braaten1999} but, like Wu, they left undetermined a parameter $g_3(\kappa)$ for the effective strength
of the three-body force near the scattering threshold \cite{Braaten1999}, which is related to Wu's parameter $\mathcal{E}_3$ \cite{Wu}.
[The \emph{difference}, $g_3(\kappa)-g_3(\kappa')$, is known for any momentum scales $\kappa$ and $\kappa'$ \cite{Braaten1999}.]

For bosons with large scattering length, $g_3(\kappa)$ was recently computed using the EFT \cite{Braaten2002}.
For other model interactions, Braaten \textit{et al} \cite{Braaten2001} used the
Monte Carlo (MC) results of the energy density \cite{Giorgini} to extract $\mathcal{E}'_3$ but, because of statistical uncertainties
of the MC data that are difficult to reduce, they did not obtain satisfying answers \cite{Braaten2001}.
In summary, $g_3(\kappa)$ remains unknown for almost \emph{all} bosonic systems.

In this paper this three-body force and a few related properties of the $\CN$-body system are studied.
Implications for dilute Bose-Einstein condensates are also explored.

We know that the interaction of \emph{two} bosons at low energy ($E\ll\hbar^2/m_\text{boson}^{}r_e^2$ and $E\ll\hbar^2/m_\text{boson}a^2$,
$r_e=$ range of interaction) is dominated by the \emph{two-body scattering length} $a$, while
$a$ is present in the small-momentum expansion of the two-boson wave function at \emph{zero} collision energy
and zero orbital angular momentum:
\begin{equation}\label{eq:adef}
\phi_\vect q=(2\pi)^3\delta(\vect q)-4\pi a/q^2+u_0+O(q^2).
\end{equation}
Analogously, the effective interaction strength of \emph{three} identical bosons at low energy should be present in
the small-$q$ expansion of the wave function of the same three bosons at \emph{zero} collision energy and zero orbital angular momentum,
$\phi^{(3)}_{\vect k_1\vect k_2\vect k_3}$. It is shown in this paper that this is indeed the case. For small momenta
$\hbar q_i\sim\hbar q$ ($\sum_{i=1}^3\vect q_i\equiv0$), it is found that 
\begin{align}
&\phi^{(3)}_{\vect q_1\vect q_2\vect q_3}=(2\pi)^6\delta(\vect q_1)\delta(\vect q_2)+G_{q_1q_2q_3}\notag\\
&\times\Big\{\Big[\sum_{i=1}^3-4\pi a(2\pi)^3\delta(\vect q_i)+32\pi^2a^2/q_i^2-16\pi^2wa^3/q_i\notag\\
&-64\pi wa^4\ln(q_i\lvert a\rvert)\Big]\!-\!D\Big\}\!+u_0\!\sum_{i=1}^3\big[(2\pi)^3\delta(\vect q_i)-8\pi a/q_i^2\big]\notag\\
&\mspace{310mu}+O(q^{-1}),\label{eq:Ddef}
\end{align}
where $G_{q_1q_2q_3}\equiv2/(q_1^2+q_2^2+q_3^2)$, and that
$D$, named \emph{three-body scattering hypervolume}
(with dimension $[\text{length}]^4$), is a suitable parameter for the three-boson interaction.\cite{note:Absa}

The definition of the three-body parameter in Eq.~\eqref{eq:Ddef} permits one to determine
this parameter in an elementary way, namely by solving the three-body Schr\"{o}dinger equation in free space and matching the solution
to Eq.~\eqref{eq:Ddef} at small momenta or its Fourier transform at large relative distances.

In Sec.~\ref{sec:asymp} of the present paper, $\phi^{(3)}_{\vect q_1\vect q_2\vect q_3}$
is expanded to the order $q^1$ at small momenta [Eqs.~\eqref{eq:phi3smallmomenta}]
and, correspondingly, its Fourier transform
expanded to the order $R^{-7}$ at large relative distances $R$ [Eqs.~\eqref{eq:phi3coordinate}].
Although the parameter $D$ first appears in the latter
expansion at the order $R^{-4}$, three higher order corrections are determined, to facilitate much more accurate determinations of $D$ from
numerical solutions to Schr\"{o}dinger equation at not-so-large relative distances.

In Sec.~\ref{sec:HS}, the results of Sec.~\ref{sec:asymp} are applied to bosons interacting through the hard-sphere (HS) potential.
By solving the 3-body Schr\"{o}dinger equation numerically the author found
\begin{equation}\label{eq:DHS}
D_\text{HS}=(1761.5430\pm0.0024)a^4.
\end{equation}

In Sec.~\ref{sec:cubic} the ground state of three identical bosons in a large periodic cubic volume of side $L$
is determined perturbatively in powers of $L^{-1}$.
The energy is expanded \cite{footnote:HuangYangExpansion} to the order $L^{-7}$ ($\hbar=m_\text{boson}=1$):
\begin{widetext}
\begin{multline}\label{eq:Enumerical}
E=\frac{12\pi a}{L^3}\bigg[1+2.837~297~479~480~619~476~67\frac{a}{L} +9.725~330~808~459~240~0570\frac{a^2}{L^2}
+\Big(-39.307~830~355~480~219~057\ln\!\frac{L}{\lvert a\rvert}\\
+95.852~723~604~821~230~29\Big)\frac{a^3}{L^3}
+\frac{3\pi a^2r_s}{L^3}
+\Big(-669.168~047~948~734~849~322\ln\!\frac{L}{\lvert a\rvert}+810.053~286~803~649~420\Big)\frac{a^4}{L^4}\\
+53.481~797~505~510~907~636\frac{a^3r_s}{L^4}\bigg]
+\frac{D}{L^6}+17.023~784~876~883~716~860\frac{aD}{L^7}+O(L^{-8}),
\end{multline}
\end{widetext}
where $r_s$ is the two-body effective range.

In Sec.~\ref{subsec:E3} Wu's parameter $\mathcal{E}_3$ \cite{Wu} is expressed in terms of $D$;
$\mathcal{E}_3$ for hard-sphere bosons is then found [Eq.~\eqref{eq:E3HS}].

In the rest of Sec.~\ref{sec:cubic} the $\CN$-boson energy and momentum distribution are found.
The importance of the parameter $u_0$ [defined in Eq.~\eqref{eq:adef}], which is \emph{absent} in the effective field theory,
is stressed. General formulas for the BEC energy and condensate depletion are obtained
[Eqs.~\eqref{eq:E0_thermodynamic} and \eqref{eq:x_thermodynamic}].

In Sec.~\ref{sec:Tmat} the scattering amplitude of three bosons at low energy is computed [Eqs.~\eqref{eq:T3}].
After a discrepancy between Ref.~\cite{Braaten1999} and our result at $r_s\ne0$ is resolved,
$g_3(\kappa)$ of Ref.~\cite{Braaten1999} is expressed in terms of the scattering hypervolume $D$
[Eq.~\eqref{eq:g3}]. The ground state energy per particle of a dilute Bose gas of hard spheres is finally determined to order $\rho^2$
[Eq.~\eqref{eq:E3pHS}].


\section{asymptotics of $\phi^{(3)}$ at small momenta or large relative distances\label{sec:asymp}}
We consider identical bosons with instantaneous interactions that are translationally, rotationally, and Galilean invariant, and finite-ranged
(\textit{ie}, limited within a finite interparticle distance $r_e$).
So the 2-body interaction $\frac{1}{2}U_{\vect k_1\vect k_2\vect k_3\vect k_4}$ conserves momentum, is invariant under
rotation or any equal shift of $\vect k_i$'s ($1\le i\le4$), and is smooth. Also, because of Bose statistics, we can symmetrize
$U$ with respect to the incoming (outgoing) momenta without losing generality \cite{footnote:Usymmetrize}:
$U_{\vect k_1\vect k_2\vect k_3\vect k_4}=U_{\vect k_2\vect k_1\vect k_3\vect k_4}=U_{\vect k_1\vect k_2\vect k_4\vect k_3}
=U^*_{\vect k_4\vect k_3\vect k_2\vect k_1}$ (the last equality is the Hermiticity condition).
Similar properties are held by the 3-body interaction,
$\frac{1}{6}U_{\vect k_1\vect k_2\vect k_3\vect k_4\vect k_5\vect k_6}$.

We will consider scattering states only. Units such that $\hbar=m_\text{boson}=1$ are used.

\subsection{Two-body special functions}
This subsection is in preparation for the rest of the paper.

We define 2-body special functions in the $l$-wave channel ($l$ is even for identical bosons),
with zero magnetic quantum number along the direction $\hat{\vect n}$:
$\phi^{(l)}_{\hat{\vect n}\vect k}$, $f^{(l)}_{\hat{\vect n}\vect k}$, $g^{(l)}_{\hat{\vect n}\vect k}$, \dots. \cite{footnote:specialfunctions}
\begin{gather}
(H\phi^{(l)}_{\hat{\vect n}})_\vect k=0,~~~~(Hf^{(l)}_{\hat{\vect n}})_\vect k=\phi^{(l)}_{\hat{\vect n}\vect k},
~~~~(Hg^{(l)}_{\hat{\vect n}})_\vect k=f^{(l)}_{\hat{\vect n}\vect k},
\label{eq:phi_f_g}\\
(HX)_\vect k\equiv k^2X_\vect k+\frac{1}{2}\int\frac{\dif^3k'}{(2\pi)^3}U_{\vect k\vect k'}X_{\vect k'},\label{eq:H}
\end{gather}
where $U_{\vect k\vect k'}\equiv U_{\vect k,-\vect k,\vect k',-\vect k'}$. For $l=0$, we write these functions simply as
$\phi_\vect k$, $f_\vect k$, $g_\vect k$, \dots. For $l\ge2$, we use usual symbols $d,g,i,\dots$ to represent $l=2,4,6\dots$.
Let
$$\phi(\vect r)\equiv\int\frac{\dif^3k}{(2\pi)^3}\phi_\vect k\e^{\I\vect k\cdot\vect r}$$
and similarly for the other functions. The amplitudes of these functions
are fixed by the following equations at $r>r_e$:
\begin{subequations}\label{eq:2bodyspecial}
\begin{align}
\phi(\vect r)&=1-a/r,\\
f(\vect r)&=-r^2/6+ar/2-ar_s/2,\\
g(\vect r)&=r^4/120-ar^3/24+ar_sr^2/12-ar_s'/24,\\
\phi^{(d)}_{\hat{\vect n}}(\vect r)&=\big(r^2/15-3a_d/r^3)P_2(\hat{\vect n}\cdot\hat{\vect r}),\\
f^{(d)}_{\hat{\vect n}}(\vect r)&=\big(-r^4/210-a_dr_dr^2/30-a_d/2r\big)P_2(\hat{\vect n}\cdot\hat{\vect r}),\\
\phi^{(g)}_{\hat{\vect n}}(\vect r)&=\big(r^4/945-105a_g/r^5\big)P_4(\hat{\vect n}\cdot\hat{\vect r}),\\
\phi^{(i)}_{\hat{\vect n}}(\vect r)&=\big(r^6/135135-10395a_i/r^7\big)P_6(\hat{\vect n}\cdot\hat{\vect r}),
\end{align}
\end{subequations}
where $P_l$ is the Legendre polynomial [$P_l(1)\equiv1$], and the $l$-wave scattering phase shift $\delta_l$ at low energy $k^2$ satisfies:
\begin{equation}\label{eq:phaseshift}
k^{2l+1}\cot\delta_l(k)=-a_l^{-1}+{r_l^{}}k^2/{2!}+{r'_l}k^4/{4!}+O(k^6),
\end{equation}
where $a_0=a$, $r_0=r_s$, $r_0'=r_s'$, $a_2=a_d$, $r_2=r_d$, $a_4=a_g$, and $a_6=a_i$.
Now define harmonic polynomials
\begin{equation}\label{eq:Qdirection}
Q^{(l)}_{\hat{\vect n}}(\vect k)\equiv k^lP_l(\hat{\vect n}\cdot\hat{\vect k}).
\end{equation}
The following more general formulas in the momentum space are directly
related to Eqs.~\eqref{eq:2bodyspecial} (by Fourier transformation):

\begin{widetext}
\begin{subequations}\label{eq:phi_f_gAllk}
\begin{align}
\phi^{(l)}_{\hat{\vect n}\vect k}&=\frac{\I^l}{(2l+1)!!}Q^{(l)}_{\hat{\vect n}}(\nabla_k)(2\pi)^3\delta(\vect k)
+\Bigl[-\frac{4\pi a_l^{}}{\I^{l}k^{2}}+\sum_{i=0}^\infty u_i^{(l)}k^{2i}\Bigr]Q^{(l)}_{\hat{\vect n}}(\vect k),\label{eq:phi_Allk}\\
f^{(l)}_{\hat{\vect n}\vect k}&=\Bigl[\frac{\nabla_k^2}{2!!(2l+3)!!}-\frac{a_l^{}r_l^{}}{2!(2l+1)!!}\Bigr]
\I^lQ^{(l)}_{\hat{\vect n}}(\nabla_k)(2\pi)^3\delta(\vect k)
+\Bigl[-\frac{4\pi a_l^{}Z}{\I^lk^4}+\sum_{i=0}^\infty \widetilde{f}_i^{(l)}k^{2i}\Bigr]Q^{(l)}_{\hat{\vect n}}(\vect k),\\
g^{(l)}_{\hat{\vect n}\vect k}&=\Bigl[\frac{\nabla_k^4}{4!!(2l+5)!!}-\frac{a_l^{}r_l^{}\nabla_k^2}{2!2!!(2l+3)!!}
-\frac{a_l^{}r_l'}{4!(2l+1)!!}\Bigr]
\I^lQ^{(l)}_{\hat{\vect n}}(\nabla_k)(2\pi)^3\delta(\vect k)
+\Bigl[-\frac{4\pi a_l^{}Z}{\I^lk^6}+\sum_{i=0}^\infty\widetilde{g}_i^{(l)}k^{2i}\Bigr]Q^{(l)}_{\hat{\vect n}}(\vect k),
\end{align}
\end{subequations}
\end{widetext}
where $\I=\sqrt{-1}\ne i$, and $Z/k^4$ and $Z/k^6$ are generalized functions
(in this paper $Z$ is merely a symbol and \emph{not} a number):
\begin{gather*}
\frac{Z}{k^4}=\frac{1}{k^4}~~(k>0),~~~\int_{\text{all }\vect k}\frac{Z}{k^4}\dif^3k=0,\\
\frac{Z}{k^6}=\frac{1}{k^6}~~(k>0),~~~\int_{\text{all }\vect k}\frac{Z}{k^6}\dif^3k=\int_{\text{all }\vect k}k^2\frac{Z}{k^6}\dif^3k=0.
\end{gather*}
$Z/k^4$ and $Z/k^6$ have $-\infty$ values at $\vect k=0$ to cancel certain integrals shown above.
They inevitably arise from the Fourier transformation of functions
such as $\lvert\vect r\rvert$. The $Z$-functions are more completely described in Appendix~\ref{sec:Z}.

The infinite series such as $\sum_{i=0}^\infty u_i^{(l)}k^{2i}$ in Eqs.~\eqref{eq:phi_f_gAllk} account for the deviations of
$\phi^{(l)}_{\hat{\vect n}}(\vect r)$, $f^{(l)}_{\hat{\vect n}}(\vect r)$, and $g^{(l)}_{\hat{\vect n}}(\vect r)$
from Eqs.~\eqref{eq:2bodyspecial} at $r<r_e$. Since $r_e<\infty$, these series are convergent.

The superscripts of $u^{(l)}_i$ and $\widetilde{f}^{(l)}_i$ will be omitted at $l=0$.

From Eqs.~\eqref{eq:phi_f_g} and \eqref{eq:phi_f_gAllk} we derive that at small $\vect k$
\begin{subequations}\label{eq:intUphi}
\begin{gather}
\frac{1}{2}\int\frac{\dif^3k'}{(2\pi)^3}U_{\vect k\vect k'}\phi^{(l)}_{\hat{\vect n}\vect k'}=
\Big[\I^{-l}4\pi a_l^{}-\sum_{i=0}^\infty u_i^{(l)}k^{2i+2}\Big]Q^{(l)}_{\hat{\vect n}}(\vect k),\\
\frac{1}{2}\int\frac{\dif^3k'}{(2\pi)^3}U_{\vect k\vect k'}f^{(l)}_{\hat{\vect n}\vect k'}
=\sum_{i=0}^\infty\big[u_i^{(l)}k^{2i}-\widetilde{f}^{(l)}_ik^{2i+2}\big]Q^{(l)}_{\hat{\vect n}}(\vect k),\\
\frac{1}{2}\int\frac{\dif^3k'}{(2\pi)^3}U_{\vect k\vect k'}g_{\vect k'}=\widetilde{f}_0+O(k^2).\label{eq:int_g}
\end{gather}
\end{subequations}

For any unknown $X_\vect k$ we have the uniqueness theorem \cite{footnote:deltak}:
\begin{equation}\label{eq:unique2}
(HX)_\vect k\equiv0\text{~(all $\vect k$)}\text{~and~}X_\vect k=o(k^{-3})~(\text{small }\vect k)\\\Rightarrow X_\vect k\equiv0.
\end{equation}

The following identity is needed in the analysis of the momentum distribution of $\CN$ particles at low density ($\Re$ stands for the real part)
\cite{footnote:a_rs_real}:
\begin{equation}\label{eq:IntPhikSquare}
\lim_{k_c\rightarrow0^+}\int_{k>k_c}\!\frac{\dif^3k}{(2\pi)^3}\Big(\lvert\phi_\vect k\rvert^2-\frac{16\pi^2a^2}{k^4}\Big)=-2\pi a^2r_s-2\Re u_0.
\end{equation}

\subsection{Asymptotics of $\phi^{(3)}_{\vect k_1\vect k_2\vect k_3}$ at small momenta}
From this point on, we
let $\vect q$'s be small momenta and $q_i$'s scale like $q^1$, while $\vect k$'s be independent from $\vect q$'s.
We will derive the following asymptotic expansions:
\begin{gather}
\phi^{(3)}_{\vect q_1\vect q_2\vect q_3}=\sum_{s=-6}^\infty T^{(s)}_{\vect q_1\vect q_2\vect q_3},\\
\phi^\vect q_\vect k\equiv\phi^{(3)}_{\vect q,-\vect q/2+\vect k,-\vect q/2-\vect k}=\sum_{s=-3}^\infty S^{(s)\vect q}_\vect k,
\end{gather}
where $T^{(s)}_{\vect q_1\vect q_2\vect q_3}$ and $S^{(s)\vect q}_\vect k$ both scale like $q^s$
(including possibly $q^s\ln^nq$, $n=1,2,\dots$). The minimum values of $s$ in the above equations will be justified below.

The 3-body Schr\"{o}dinger equation can be written in two special forms \cite{footnote:phi3}:
\begin{gather}
G_{q_1q_2q_3}^{-1}\phi^{(3)}_{\vect q_1\vect q_2\vect q_3}\!=\!
-\big(\sum_{i=1}^3\frac{1}{2}\int_{\vect k'}U_{\vect p_i\vect k'}\phi^{\vect q_i}_{\vect k'}\big)-U^{\phi}_{\vect q_1\vect q_2\vect q_3},
\label{eq:Schrodinger3qqq}\\
(H\phi^\vect q)_\vect k+{3}q^2\phi^\vect q_\vect k/{4}+W^\vect q_\vect k=0,\label{eq:Schrodinger3kq}
\end{gather}
where $G_{q_1q_2q_3}=2/(q_1^2+q_2^2+q_3^2)$,
\begin{gather}
\vect p_1=(\vect q_2-\vect q_3)/2,\text{ and similarly for $\vect p_2$, $\vect p_3$},\label{eq:p}\\
U^{\phi}_{\vect q_1\vect q_2\vect q_3}=\frac{1}{6}\int_{\vect k_1'\vect k_2'}U_{\vect q_1\vect q_2\vect q_3\vect k_1'\vect k_2'\vect k_3'}
\phi^{(3)}_{\vect k_1'\vect k_2'\vect k_3'},\\
W^\vect q_\vect k=\Big[\frac{1}{2}\int_{\vect k'}U_{-\vect q/2+\vect k,\vect q\vect k'\vect k''}
\phi^{(3)}_{-\vect q/2-\vect k,\vect k'\vect k''}+(\vect q\leftrightarrow-\vect q)\Big]\notag\\
~~~~~~~~~~~~~~~~~~~~~~~~~~~~~~~~~~~~~~~~~~~~
+U^{\phi}_{-\vect q/2+\vect k,-\vect q/2-\vect k,\vect q},
\end{gather}
$\int_{\vect k'}$ is the shorthand for $\int\frac{\dif^3k'}{(2\pi)^3}$, and
$\int_{\vect k'\vect k''}$ stands for $\int\int\frac{\dif^3k'}{(2\pi)^3}\frac{\dif^3k''}{(2\pi)^3}$.

At small $\vect q$'s we have Taylor expansions:
\begin{subequations}
\begin{gather}
U^\phi_{\vect q_1\vect q_2\vect q_3}=\kappa_0+\kappa_1(q_1^2+q_2^2+q_3^2)+O(q^4),\label{eq:UphiTaylor}\\
W^\vect q_\vect k=\sum_{s=0,2,4\dots}q^sW^{(s)}_{\hat{\vect q}\vect k},~~~~~W^{(0)}_{\hat{\vect q}\vect k}\equiv W^{(0)}_\vect k.\label{eq:WTaylor}
\end{gather}
\end{subequations}

Now we fix the overall amplitude of $\phi^{(3)}$ ($\sum_{i=1}^3\vect q_i\equiv0$):
$
\phi^{(3)}_{\vect q_1\vect q_2\vect q_3}\equiv(2\pi)^6\delta(\vect q_1)\delta(\vect q_2)+\text{higher order terms}.
$
Because $\delta(\lambda\vect q)=\lambda^{-3}\delta(\vect q)$, $\delta(\vect q)$ scales like $q^{-3}$.
So $s\ge-6$ for $T^{(s)}$, and
\begin{equation}
T^{(-6)}_{\vect q_1\vect q_2\vect q_3}=(2\pi)^6\delta(\vect q_1)\delta(\vect q_2).
\end{equation}
So $T^{(-6)}_{\vect q,-\vect q/2+\vect k,-\vect q/2-\vect k}=(2\pi)^3\delta(\vect q)(2\pi)^3\delta(\vect k)\sim q^{-3}k^{-3}$,
indicating that $s\ge-3$ for $S^{(s)\vect q}_\vect k$.

The following statement is now true at $s_1=-6$:

\textbf{Statement $\mathbf{s_1}$}: all the functions $T^{(s)}$ for $s\le s_1$,
and all the $S^{(s)}$ for $s\le s_1+2$, have been formally determined.

We can then do the following expansions at small $\vect q$,
for $-6\le s\le s_1$:
\begin{equation}\label{eq:tdef}
T^{(s)}_{\vect q,-\vect q/2+\vect k,-\vect q/2-\vect k}=\sum_nt^{(n,s-n)}_{\vect q,\vect k},
\end{equation}
where $t^{(n,s-n)}_{\vect q,\vect k}$ scales like $q^nk^{s-n}$.
Note also that
$S^{(-3)\vect q}_\vect k+S^{(-2)\vect q}_\vect k+\cdots=T^{(-6)}_{\vect q,-\vect q/2+\vect k,-\vect q/2-\vect k}
+T^{(-5)}_{\vect q,-\vect q/2+\vect k,-\vect q/2-\vect k}+\cdots$ [$=\phi^{(3)}_{\vect q,-\vect q/2+\vect k,-\vect q/2-\vect k}$].
Therefore, the asymptotic expansion of $S^{(s_1+3)\vect q}_\vect k$ at small $\vect k$ has been determined to the order $k^{-3}$:
\begin{subequations}\label{eq:S_eq}
\begin{equation}\label{eq:S_eq1}
S^{(s_1+3)\vect q}_\vect k=\sum_{m=-s_1-9}^{-3}t^{(s_1+3,m)}_{\vect q,\vect k}+O(q^{s_1+3}k^{-2}).
\end{equation}
Equations~\eqref{eq:tdef} and \eqref{eq:S_eq1} ensure the continuity of the wave function $\phi^{(3)}$
across two connected regions.
Extracting all the terms that scale like $q^{s_1+3}$ from Eq.~\eqref{eq:Schrodinger3kq}, we get
\begin{equation}\label{eq:S_eq2}
(HS^{(s_1+3)\vect q})_\vect k=-{3}q^2S^{(s_1+1)\vect q}_\vect k/{4}-W^{(s_1+3)}_{\hat{\vect q}\vect k}q^{s_1+3}.
\end{equation}
\end{subequations}
If $W^{(s_1+3)}_{\hat{\vect q}\vect k}\ne0$ (\textit{ie}, if $s_1+3=0,2,4\cdots$), we take it as formal input.
Solving Eqs.~\eqref{eq:S_eq}, with the help of \eqref{eq:phi_f_g} and \eqref{eq:phi_f_gAllk},
we thus determine $S^{(s_1+3)\vect q}_\vect k$. The uniqueness of the solution is guaranteed by \eqref{eq:unique2}.
 
Once $S^{(s_1+3)}$ is determined, the right hand side of Eq.~\eqref{eq:Schrodinger3qqq} can be determined
up to the order $q^{s_1+3}$, if the coefficients of the Taylor expansion of $U^\phi_{\vect q_1\vect q_2\vect q_3}$ [see Eq.~\eqref{eq:UphiTaylor}]
are regarded as formal input. Solving Eq.~\eqref{eq:Schrodinger3qqq}, and noting that $G_{q_1q_2q_3}^{-1}\sim q^2$, we thus
determine $T^{(s_1+1)}$. 

So now the truth of Statement $(s_1+1)$ is established.

We can thus formally determine all the functions $T^{(s)}$ and $S^{(s)}$,
by repeating the above routine, starting from $s_1=-6$. The results of this program are shown below.

\textbf{Step 1.}
$S^{(-3)\vect q}_\vect k=(2\pi)^3\delta(\vect q)(2\pi)^3\delta(\vect k)+O(q^{-3}k^{-2})$ at small $\vect k$,
and $(HS^{(-3)\vect q})_\vect k=0$, so
\begin{equation}
S^{(-3)\vect q}_\vect k=(2\pi)^3\delta(\vect q)\phi_\vect k.
\end{equation}

\textbf{Step 2.}
With the help of Eqs.~\eqref{eq:intUphi}, and noting that $\sum_{i=1}^3\vect q_i\equiv0$, we get
\begin{equation}
T^{(-5)}_{\vect q_1\vect q_2\vect q_3}=\sum_{i=1}^3-(4\pi a/p_i^2)(2\pi)^3\delta(\vect q_i).
\end{equation}
Expanding $T^{(-5)}_{\vect q,-\vect q/2+\vect k,-\vect q/2-\vect k}$ at small $\vect q$, we get all the functions $t^{(n,m)}_{\vect q\vect k}$ for $n+m=-5$;
shown below are those in the range $-2\le n\le3$ (zeros are omitted):
\begin{subequations}
\begin{align}
t^{(-2,-3)}_{\vect q\vect k}&=-(8\pi a/q^2)(2\pi)^3\delta(\vect k),\\
t^{(0,-5)}_{\vect q\vect k}&=-\pi a(\hat{\vect q}\cdot\nabla_k)^2(2\pi)^3\delta(\vect k),\\
t^{(2,-7)}_{\vect q\vect k}&=-(\pi a/48)(\hat{\vect q}\cdot\nabla_\vect k)^4(2\pi)^3\delta(\vect k)q^2.
\end{align}
\end{subequations}

\textbf{Step 3.}
$S^{(-2)\vect q}_\vect k=t^{(-2,-3)}_{\vect q\vect k}+O(q^{-2}k^{-2})$ at small $\vect k$, and $(HS^{(-2)\vect q})_\vect k=0$, so
\begin{equation}
S^{(-2)\vect q}_\vect k=-(8\pi a/q^2)\phi_\vect k.
\end{equation}

\textbf{Step 4.} Using the same method as in Step 2, we get
\begin{equation}
T^{(-4)}_{\vect q_1\vect q_2\vect q_3}=32\pi^2a^2G_{q_1q_2q_3}\sum_{i=1}^{3}q_i^{-2}.
\end{equation}
At small $\vect q$ we have the following expansions (named ``$Z$-$\delta$ expansions"; see Appendix~\ref{sec:Z-delta} for details):
\begin{subequations}\label{eq:Z-deltaExample1}
\begin{align}
&(k^2+3q^2/4)^{-1}=k^{-2}-\sqrt{3}(2\pi)^3\delta(\vect k)q/8\pi-3q^2Z/4k^4\notag\\
&\quad\quad\quad\quad+{\sqrt{3}}\nabla_k^2(2\pi)^3\delta(\vect k)q^3/{64\pi}+9q^4Z/16k^6\notag\\
&\quad\quad\quad\quad-{3\sqrt{3}}\nabla_k^4(2\pi)^3\delta(\vect k)q^5/{5120\pi}+O(q^6),\label{eq:Z-deltaExample1a}
\end{align}
\begin{align}
&\big(\lvert\vect k+\vect q/2\rvert^{-2}\!+\!\lvert\vect k-\vect q/2\rvert^{-2}\big)/(k^2\!+\!3q^2/4)
\!=\!(2\pi)^3\delta(\vect k)/6q\notag\\&\quad
+2Z/k^4+\big[-\nabla_k^2/48+\big(1/24-\sqrt{3}/16\pi\big)Q^{(d)}_{\hat{\vect q}}(\nabla_k)\big]\notag\\&\quad\times(2\pi)^3\delta(\vect k)q
+\big[4ZQ^{(d)}_{\hat{\vect q}}(\vect k)/3k^8-4Z/3k^6\big]q^2\notag\\&\quad
+\big[\nabla_k^4/{1280}-\big(1/448-3\sqrt{3}/896\pi\big)\nabla_k^2Q^{(d)}_{\hat{\vect q}}(\nabla_k)\notag\\&\quad
+\big(19/10080-3\sqrt{3}/896\pi\big)Q^{(g)}_{\hat{\vect q}}(\nabla_k)
\big](2\pi)^3\delta(\vect k)q^3\notag\\
&\quad\quad\quad\quad\quad\quad\quad\quad\quad\quad\quad\quad\quad\quad\quad\quad
+O(q^4),\label{eq:Z-deltaExample1b}
\end{align}
\end{subequations}
so $T^{(-4)}_{\vect q,-\vect q/2+\vect k,-\vect q/2-\vect k}=\sum_{n=-2}^\infty t^{(n,-4-n)}_{\vect q,\vect k}$, where
\begin{subequations}\begin{align}
t^{(-2,-2)}_{\vect q,\vect k}&=32\pi^2a^2/q^2k^2,\\
t^{(-1,-3)}_{\vect q,\vect k}&=4\pi wa^2(2\pi)^3\delta(\vect k)/q,\\
t^{(0,-4)}_{\vect q,\vect k}&=40\pi^2a^2Z/k^4,
\end{align}\end{subequations}
and for brevity higher order terms [which can readily be obtained from Eqs.~\eqref{eq:Z-deltaExample1}] are not shown.

\textbf{Step 5.}
$S^{(-1)\vect q}_\vect k=t^{(-1,-3)}_{\vect q,\vect k}+O(q^{-1}k^{-2})$ at small $\vect k$, and $(HS^{(-1)\vect q})_\vect k=0$, so
\begin{equation}
S^{(-1)\vect q}_{\vect k}=(4\pi wa^2/q)\phi_\vect k.
\end{equation}

\textbf{Step 6.} This is similar to Steps 2 and 4.
\begin{equation}
T^{(-3)}_{\vect q_1\vect q_2\vect q_3}=-16\pi^2wa^3G_{q_1q_2q_3}\sum\limits_{i=1}^{3}q_i^{-1}
+u_0\sum_{i=1}^3(2\pi)^3\delta(\vect q_i).
\end{equation}
Doing the $Z$-$\delta$ expansion (Appendix~\ref{sec:Z-delta}), one finds that
$T^{(-3)}_{\vect q,-\vect q/2+\vect k,-\vect q/2-\vect k}=\sum_{n=-3}^\infty t^{(n,-3-n)}_{\vect q,\vect k}$.
For brevity only one of the terms is shown here (to be used in the next step):
\begin{align}
t^{(0,-3)}_{\vect q,\vect k}&=\big[16wa^3\ln(q\lvert a\rvert)+2u_0+(14\pi/\sqrt{3}-16)wa^3\big]\notag\\
&\quad\quad\,\,\times(2\pi)^3\delta(\vect k)-32\pi^2wa^3Z_{1/\lvert a\rvert}(k)/k^3.\label{eq:t0-3}
\end{align}
This equation remains \emph{unaffected} if both $\lvert a\rvert$'s in the logarithm and in the subscript of $Z$
are replaced by any other length scale simultaneously, because of Eq.~\eqref{eq:Z_Shiftb}.

\textbf{Step 7.}
Substituting $s_1=-3$ into Eq.\eqref{eq:S_eq2}, we get
$
(HS^{(0)\vect q})_\vect k=6\pi a\phi_\vect k-W^{(0)}_\vect k.
$
Because the right hand side does not depend on $\hat{\vect k}$ or $\vect q$, we can introduce
a single function $d_\vect k$, which is independent of $\hat{\vect k}$ and satisfies
\begin{subequations}\label{eq:d}
\begin{equation}\label{eq:d1}
(Hd)_\vect k=6\pi a\phi_\vect k-W^{(0)}_\vect k,
\end{equation}
and get $S^{(0)\vect q}_\vect k=d_\vect k+(\text{linear combination of $\phi^{(l)}_{\hat{\vect q}\vect k}$})$.
Equation~\eqref{eq:d1} does not completely determine $d_\vect k$, since $d_\vect k+\eta\phi_\vect k$ satisfies the same
equation. At small $\vect k$ [Eq.~\eqref{eq:S_eq1}],
\begin{multline*}
S^{(0)\vect q}_\vect k=\big[-{2}\pi aQ^{(d)}_{\hat{\vect q}}(\nabla_k)/{3}+16wa^3\ln(q\lvert a\rvert)\big](2\pi)^3\delta(\vect k)\\
+d_a(\vect k)+O(k^{-2}),
\end{multline*}
where
\begin{align}
d_a(\vect k)&\equiv-(\pi a/3)\nabla_k^2(2\pi)^3\delta(\vect k)+40\pi^2a^2Z/k^4\notag\\
&\quad\,+\big[2u_0+(14\pi/\sqrt{3}-16)wa^3\big](2\pi)^3\delta(\vect k)\notag\\
&\quad\,-32\pi^2wa^3Z_{1/\lvert a\rvert}(k)/k^3
\end{align}
does not depend on $\hat{\vect k}$ or $\vect q$. Noting Eq.~\eqref{eq:phi_Allk}, we can now complete
our defintion of $d_{\vect k}$ \cite{footnote:dk} and determine $S^{(0)\vect q}_\vect k$:
\begin{equation}\label{eq:d2}
d_{\vect k}=d_a(\vect k)+O(k^{-2})~~~\text{at small $\vect k$}.
\end{equation}
\end{subequations}
\begin{equation}\label{eq:S0}
S^{(0)\vect q}_\vect k=16wa^3\ln(q\lvert a\rvert)\phi_\vect k+10\pi a\phi^{(d)}_{\hat{\vect q}\vect k}+d_\vect k.
\end{equation}

\textbf{Step 8.} This is similar to Steps 2, 4, and 6. Extracting all
the terms that scale like $q^0$ from both sides of Eq.~\eqref{eq:Schrodinger3qqq}, and solving the resultant equation,
we get
\begin{align}
&T^{(-2)}_{\vect q_1\vect q_2\vect q_3}=G_{q_1q_2q_3}\big[-64\pi wa^4\ln(q_1q_2q_3\lvert a\rvert^3)-D\big]\notag\\
&\quad\quad\quad\quad\,\,-\sum_{i=1}^38\pi au_0/q_i^2,\\
&D\!\equiv\!\frac{3}{2}\int_{\vect k'}\!U_{0,\vect k'}d_{\vect k'}+\frac{1}{6}\int_{\vect k_1'\vect k_2'}\!\!\!
U_{0,0,0,\vect k_1'\vect k_2'\vect k_3'}\phi^{(3)}_{\vect k_1'\vect k_2'\vect k_3'}\!-\!18\pi au_0.\label{eq:D_complicated}
\end{align}

The subsequent steps are similar to the above ones but considerably lengthier; details will not be shown.
At the completion of Step 14 the following results are accumulated:
\begin{widetext}
\begin{subequations}\label{eq:phi3smallmomenta}
\begin{multline}\label{eq:phi3qk_res}
\phi^{(3)}_{-\vect q/2+\vect k,-\vect q/2-\vect k,\vect q}=\Bigl[(2\pi)^3\delta(\vect q)-\frac{8\pi a}{q^2}+\frac{4\pi wa^2}{q}
+16wa^3\ln(q\lvert a\rvert)+24\sqrt{3}wa^4q\ln(q\lvert a\rvert)+\xi_1q
-\frac{32\sqrt{3}wa^5}{\pi}q^2\ln^2(q\lvert a\rvert)\\
-\xi_2q^2\ln(q\lvert a\rvert)-\zeta_3q^3\ln(q\lvert a\rvert)+\xi_3q^3\Bigr]\phi_\vect k+\Bigl[-3\pi wa^2q-12wa^3q^2\ln(q\lvert a\rvert)
-18\sqrt{3}wa^4q^3\ln(q\lvert a\rvert)-\frac{3\xi_1}{4}q^3\Bigr]f_\vect k+\frac{9\pi w}{4}a^2q^3g_\vect k\\
+\Bigl[10\pi a-10\pi\bigl(2\pi-3\sqrt{3}\bigr)a^2q-4wa^3q^2\ln(q\lvert a\rvert)+\xi_3^{(d)}q^3\Bigr]\phi^{(d)}_{\hat{\vect q}\vect k}
+\frac{15\pi}{2}\bigl(2\pi-3\sqrt{3}\bigr)a^2q^3f^{(d)}_{\hat{\vect q}\vect k}\\
+\Bigl[-\frac{9\pi}{2}aq^2+\frac{3\pi}{4}\bigl(76\pi-135\sqrt{3}\bigr)a^2q^3\Bigr]\phi^{(g)}_{\hat{\vect q}\vect k}
+d_\vect k+q^2d^{(2)}_{\hat{\vect q}\vect k}+O(q^4),
\end{multline}
\begin{multline}\label{eq:phi3qqq_res}
\phi^{(3)}_{\vect q_1\vect q_2\vect q_3}=(2\pi)^3\delta(\vect q_1)(2\pi)^3\delta(\vect q_2)+\frac{2}{q_1^2+q_2^2+q_3^2}\sum_{i=1}^3\Big[
-4\pi a(2\pi)^3\delta(\vect q_i)+\frac{32\pi^2a^2}{q_i^2}-\frac{16\pi^2wa^3}{q_i}-64\pi wa^4\ln(q_i\lvert a\rvert)-\frac{D}{3}\\
-96\sqrt{3}\,\pi wa^5q_i\ln(q_i\lvert a\rvert)-4\pi a\xi_1q_i+128\sqrt{3}wa^6q_i^2\ln^2(q_i\lvert a\rvert)+4\pi a\xi_2q_i^2\ln(q_i\lvert a\rvert)
+40\pi^2aa_dQ^{(d)}_{\hat{\vect q}_i}(\vect p_i)+4\pi a\zeta_3q_i^3\ln(q_i\lvert a\rvert)\\
-4\pi a\xi_3q_i^3-40\pi^2\big(2\pi-3\sqrt{3}\big)a^2a_dq_iQ^{(d)}_{\hat{\vect q}_i}(\vect p_i)\Big]
+\sum_{i=1}^3\Big[(2\pi)^3\delta(\vect q_i)\big(u_0+u_1p_i^2+u_2p_i^4\big)-\frac{8\pi a}{q_i^2}\big(u_0+u_1p_i^2\big)
+\frac{4\pi wa^2}{q_i}\big(u_0+u_1p_i^2\big)\\
+16wa^3\ln(q_i\lvert a\rvert)u_0+24\sqrt{3}wa^4q_i\ln(q_i\lvert a\rvert)u_0+\xi_1q_iu_0-3\pi wa^2q_i\widetilde{f}_0\Big]
+\chi_0+O(q^2),
\end{multline}
\end{subequations}
\end{widetext}
where $q^2d^{(2)}_{\hat{\vect q}\vect k}$ is a quadratic \emph{polynomial} of $\vect q$
[and for any rotation $\mathrm{r}$, $d^{(2)}_{\mathrm{r}\hat{\vect q},\mathrm{r}\vect k}=d^{(2)}_{\hat{\vect q}\vect k}$], and
\begin{subequations}
\begin{equation}
\xi_1\equiv\sqrt{3}D/8\pi-8\big(\sqrt{3}-\pi/3\big)wa^4-3\pi wa^3r_s/2,
\end{equation}
\begin{equation}
\xi_2\equiv\! aD/\sqrt{3}\,\pi^2\!-\!\big(260/9+128/\sqrt{3}\,\pi\big)wa^5+2wa^4r_s,\\
\end{equation}
\begin{equation}
\xi_3^{(d)}\equiv{5}\bigl(9\sqrt{3}-4\pi\bigr)wa^4/3+15\pi\bigl(2\pi-3\sqrt{3}\bigr)a^2a_dr_d/4,
\end{equation}
\begin{equation}
\zeta_3\equiv\big(24/\pi+16/\sqrt{3}\big)wa^6+9\sqrt{3}wa^5r_s,
\end{equation}
\begin{align}
\xi_3&\equiv-\big[\big(1/8\pi^2+1/12\sqrt{3}\,\pi\big)a+3\sqrt{3}\,r_s/64\pi\big]aD\notag\\
&\quad+\big({17}\sqrt{3}/2+{22}/{\pi}+{353\pi}/{27}\big)wa^6\notag\\
&\quad+\big({11\sqrt{3}}/{4}-{7\pi}/{6}\big)wa^5r_s+{9}\pi wa^4r_s^2/16\notag\\
&\quad+{3}\pi wa^3r_s'/32+\big(10\pi^2-{45}\sqrt{3}\,\pi/4\big)aa_d,
\end{align}
\end{subequations}
\begin{equation}
\chi_0\equiv9\pi au_1-3\omega_1/2-2\kappa_1-\int_{\vect k'}U_{0\vect k'}d^{(2)}_{\hat{\vect q}\vect k'}.
\end{equation}
$\kappa_1$ is defined in Eq.~\eqref{eq:UphiTaylor}, $\int_{\vect k'}\equiv\int\frac{\dif^3k'}{(2\pi)^3}$,
and $\omega_1$ is a coefficient in the following Taylor expansion at small $\vect k$:
\begin{equation}
\frac{1}{2}\int_{\vect k'}U_{\vect k\vect k'}d_{\vect k'}=\omega_0+\omega_1k^2+O(k^4).
\end{equation}
The other symbols in Eqs.~\eqref{eq:phi3smallmomenta} are defined previously:
$a$, $r_s$, $r_s'$, $a_d$, and $r_d$ in Eq.~\eqref{eq:phaseshift},
$\phi_\vect k$, $f_\vect k$, $g_\vect k$, $\phi^{(d)}_{\hat{\vect q}\vect k}$, $f^{(d)}_{\hat{\vect q}\vect k}$, and
$\phi^{(g)}_{\hat{\vect q}\vect k}$ in Eqs.~\eqref{eq:phi_f_g} and \eqref{eq:phi_f_gAllk},
$u_i$ and $\widetilde{f}_0$ in Eqs.~\eqref{eq:phi_f_gAllk} with $l=0$, $w\equiv4\pi/3-\sqrt{3}$,
$\sum_{i=1}^3\vect q_i\equiv0$, $\vect p_i$ in Eq.~\eqref{eq:p}, $Q^{(d)}$ in Eq.~\eqref{eq:Qdirection}
with $l=2$, and $d_\vect k$ in Eqs.~\eqref{eq:d}.

At the order $q^2$ in the expansion of $\phi^{(3)}_{\vect q_1\vect q_2\vect q_3}$, one will encounter another 3-body parameter, $D'$,
in the term $-D'G_{q_1q_2q_3}\sum_{i=1}^4q_i^4$. [Another contribution at the same order, $\chi_1(q_1^2+q_2^2+q_3^2)$, is smooth and
contributes nothing to the Fourier-transformed wave function at large relative distances.]
$D'$ is independent from $D$ for general interactions.

In general, if one expands $\phi^{(3)}_{\vect q_1\vect q_2\vect q_3}$ to the order $q^m$, one will encounter a total of exactly
\begin{equation}
N_3(m)=\mathrm{round}\big(m'^2/48+m'/3+89/72\big)
\end{equation}
independent 3-body parameters that contribute to the Fourier transform of $\phi^{(3)}_{\vect q_1\vect q_2\vect q_3}$
at large relative distances, where $m'=m$ (if $m$ is even), $m'=m-1$ (if $m$ is odd), and ``$\mathrm{round}$" rounds to the nearest integer.

\subsection{Asymptotics of $\phi^{(3)}(\vect r_1\vect r_2\vect r_3)$ at large relative distances}
$$\phi^{(3)}(\vect r_1\vect r_2\vect r_3)\equiv\int\frac{\dif^3k_1}{(2\pi)^3}\frac{\dif^3k_2}{(2\pi)^3}\phi^{(3)}_{\vect k_1\vect k_2\vect k_3}
\exp\Big(\sum_{i=1}^3\I\vect k_i\cdot\vect r_i\Big),$$
where $\sum_{i=1}^3{\vect k_i}\equiv0$. We consider the asymptotic expansions of $\phi^{(3)}(\vect r_1\vect r_2\vect r_3)$
in two different limits: 1) the distance $r$ between two bosons is fixed, but the distance between their center of mass
and the third boson, $R$, is large, or 2) all three interparticle distances, $s_1$, $s_2$, and $s_3$, are large but their ratio is fixed.
These two expansions are respectively
\begin{widetext}
\begin{subequations}\label{eq:phi3coordinate}
\begin{multline}\label{eq:phi3kq_res_coordinate_final}
\phi^{(3)}(\vect r/2,-\vect r/2,\vect R)=\bigg[1-\frac{2a}{R}+\frac{2wa^2}{\pi R^2}-\frac{4wa^3}{\pi R^3}
+\frac{24\sqrt{3}wa^4(\tau-3/2)-\xi_1}{\pi^2R^4}
+\frac{32\sqrt{3}wa^5(6\tau-11)-3\pi\xi_2}{2\pi^2R^5}\\
+\frac{\zeta_3(12\tau-25)+12\xi_3}{\pi^2R^6}+\frac{\zeta_4(60\tau-137)+30\xi_4}{\pi R^7}\bigg]\phi(\vect r)
+\bigg[\frac{3wa^2}{\pi R^4}-\frac{18wa^3}{\pi R^5}
+\frac{18\sqrt{3}wa^4(12\tau-25)-9\xi_1}{\pi^2R^6}\\
+\frac{48\sqrt{3}wa^5(60\tau-137)-45\pi\xi_2}{2\pi^2R^7}\bigg]f(\vect r)
+\bigg(\frac{27wa^2}{\pi R^6}-\frac{270wa^3}{\pi R^7}\bigg)g(\vect r)
+\bigg[-\frac{15a}{2R^3}+\frac{40(2\pi-3\sqrt{3})a^2}{\pi R^4}-\frac{15wa^3}{\pi R^5}+\frac{24\xi_3^{(d)}}{\pi^2R^6}\\
-\frac{12\sqrt{3}wa^5(210\tau-457)}{7\pi^2R^7}+\frac{105\xi_4^{(d)}}{2\pi R^7}
\bigg]\phi^{(d)}_{\hat{\vect R}}(\vect r)
+\bigg[\frac{180(2\pi-3\sqrt{3})a^2}{\pi R^6}-\frac{315wa^3}{2\pi R^7}\bigg]f^{(d)}_{\hat{\vect R}}(\vect r)\\
+\bigg[-\frac{945a}{8R^5}+\frac{144(76\pi-135\sqrt{3})a^2}{\pi R^6}-\frac{945wa^3}{4\pi R^7}
\bigg]\phi^{(g)}_{\hat{\vect R}}(\vect r)
-\frac{135135a}{32R^7}\phi^{(i)}_{\hat{\vect R}}(\vect r)
+O(R^{-8}),
\end{multline}
\begin{multline}\label{eq:phi3qqq_res_coordinate}
\phi^{(3)}(\vect r_1,\vect r_2,\vect r_3)=1+\bigg[\sum_{i=1}^3-\frac{a}{s_i}+\frac{4a^2\theta_i}{\pi R_is_i}
-\frac{2wa^3}{\pi B^2s_i}+\frac{8\sqrt{3}wa^4(t-1-\theta_i\cot2\theta_i)}{\pi^2B^4}\bigg]-\frac{\sqrt{3}D}{8\pi^3B^4}
\\+\sum_{i=1}^3\bigg\{\frac{36wa^5\big[(2t-3)\sin3\theta_i-2\theta_i\cos3\theta_i\big]-\sqrt{3}a\xi_1\sin3\theta_i}{\pi^2B^5\sin2\theta_i}
-\frac{96wa^6\big[3\theta_i^2\sin4\theta_i+(6t-11)\theta_i\cos4\theta_i\big]}{\pi^3B^6\sin2\theta_i}\\
+\frac{3\sqrt{3}a\xi_2\theta_i\cos4\theta_i}{\pi^2B^6\sin2\theta_i}
+\frac{45\sqrt{3}aa_d\big(24\theta_i-8\sin4\theta_i+\sin8\theta_i\big)}{8\pi B^6\sin^32\theta_i}P_2(\hat{\vect R}_i\cdot\hat{\vect s}_i)
+\frac{\sqrt{3}a\zeta_3\big[12\theta_i\cos5\theta_i+(25-12t)\sin5\theta_i\big]}{\pi^2B^7\sin2\theta_i}\\
-\frac{12\sqrt{3}a\xi_3\sin5\theta_i}{\pi^2B^7\sin2\theta_i}-45\Big(2\sqrt{3}
-\frac{9}{\pi}\Big)\frac{a^2a_d\sin^2\theta_i\big(9+10\cos2\theta_i+2\cos4\theta_i\big)}{B^7\cos^3_{}\theta_i}P_2(\hat{\vect R}_i\cdot\hat{\vect s}_i)
\bigg\}+O(B^{-8}),
\end{multline}
\end{subequations}
\end{widetext}
where $\tau\equiv\ln(\gt R/\lvert a\rvert)$, $t\equiv\ln(\gt B/\lvert a\rvert)$, $\gt\equiv\e^\gamma=1.78107\cdots$,
$\gamma$ is Euler's constant, and
\begin{gather}
\vect s_1=\vect r_2-\vect r_3,~~~\vect s_2=\vect r_3-\vect r_1,~~~\vect s_3=\vect r_1-\vect r_2,\label{eq:s}\\
\vect R_1=\vect r_1-(\vect r_2+\vect r_3)/2,\text{ and similarly for $\vect R_2$ and $\vect R_3$},\notag\\
B=\sqrt{(s_1^2+s_2^2+s_3^2)/2},~~~\theta_i=\arctan(2R_i/\sqrt{3}\,s_i).\notag
\end{gather}
\begin{subequations}
\begin{equation}
\zeta_4\equiv42\sqrt{3}\,wa^6r_s/5\pi-\big(48/5\pi^2+32/5\sqrt{3}\,\pi\big)wa^7,
\end{equation}
\begin{align}
\xi_4&\equiv\big[\big({1}/{10\pi^3}+{1}/{15\sqrt{3}\,\pi^2}\big)a-{7\sqrt{3}\,r_s}/{80\pi^2}\big]a^2D\notag\\
&\quad-\big({536}/{25\pi^2}+{124}/{27}-{86}/{25\sqrt{3}\,\pi}\big)wa^7\notag\\
&\quad+\big({65}/{3}+{309\sqrt{3}}/{25\pi}\big)wa^6r_s-6wa^5r_s^2/5\notag\\
&\quad-\big(60\pi-87\sqrt{3}\big)a^2a_d-9wa^4r_s'/20,
\end{align}
\begin{align}
\xi_4^{(d)}&\equiv{\sqrt{3}\,aD}/{28\pi^2}-\big({416}/{21}-{8384\sqrt{3}}/{245\pi}\big)wa^5\notag\\
&\quad-3wa^4r_s/7-3wa^3a_dr_d/2.
\end{align}
\end{subequations}
$R_i=B\sin\theta_i$, $s_i=\frac{2}{\sqrt{3}}B\cos\theta_i$, and $\sum_{i=1}^3\cos2\theta_i\equiv0$.

Equation \eqref{eq:phi3qqq_res_coordinate} derives from \eqref{eq:phi3qqq_res}. 

The terms up to the order $R^{-6}$ on the right hand side of Eq.~\eqref{eq:phi3kq_res_coordinate_final} derive from
\eqref{eq:phi3qk_res}; those of the order $R^{-7}$, however, are inferred from Eq.~\eqref{eq:phi3qqq_res_coordinate} and
the continuity of $\phi^{(3)}(\vect r/2,-\vect r/2,\vect R)$
across two connected regions: $r\sim O(r_e)$ and $r\sim O(R)$ (they join at $r_e\ll r\ll R$), and also the Schr\"{o}dinger equation at large $R$
\begin{equation}
\big(\widetilde{H}-3\nabla_R^2/4\big)\phi^{(3)}(\vect r/2,-\vect r/2,\vect R)=0,
\end{equation}
where $\widetilde{H}$ is the coordinate representation of the hamiltonian $H$ for two-body relative motion
[Eq.~\eqref{eq:H}].

\section{Low-energy effective interaction of 3 hard-sphere bosons\label{sec:HS}}
In this section, we consider the hard-sphere (HS) interaction. The 2-body potential $V(r)=0$ ($r>1$),
$V(r)=+\infty$ ($r<1$) ($a=1$ in this section), and there is no 3-body potential.
We numerically solve Schr\"{o}dinger equation at zero energy in the coordinate representation,
in conjunction with Eq.~\eqref{eq:phi3kq_res_coordinate_final} at large $R$, to determine $D$.\cite{footnote:HS}

For 3-body configurations excluded by the repulsive interaction, $\phi^{(3)}$ vanishes;
for allowed configurations, $\phi^{(3)}$ satisfies the free Schr\"{o}dinger equation.
So $\sum_{i=1}^3\nabla_i^2\phi^{(3)}(\vect r_1,\vect r_2,\vect r_3)$ is nonzero on the boundary $\mathcal{B}$ between these two regions only
(in fact, it has a $\delta$-function singularity on $\mathcal{B}$). Thus
\begin{align}\label{eq:phi3HS}
\phi^{(3)}(\vect r_1\vect r_2\vect r_3)&=1-\frac{4\sqrt{3}}{\pi}\!\!\int_{-1}^1\dif c'\int_{R_{\min}(c')}^\infty\!\!\!F(R',c')R'^2\dif R'\notag\\
\times\Big\{&\mathcal{G}\big[s_1,s_2,s_3;1,s_-(R',c'),s_+(R',c')\big]\notag\\
+&\mathcal{G}\big[s_1,s_2,s_3;s_-(R',c'),1,s_+(R',c')\big]\notag\\
+&\mathcal{G}\big[s_1,s_2,s_3;s_-(R',c'),s_+(R',c'),1\big]\Big\},
\end{align}
for some function $F(R',c')$,
where $s_i$'s are defined in Eq.~\eqref{eq:s},
$c'$ is the cosine of the angle formed by the line connecting two bosons with distance $1$ and the line connecting
their center-of-mass to the third boson,
\begin{subequations}
\begin{gather}
R_{\min}(c')=\big(\lvert c'\rvert+\sqrt{c'^2+3}\big)/2,\\
s_{\pm}(R',c')=\sqrt{1/4\pm R'c'+R'^2},
\end{gather}
\end{subequations}
and
\begin{subequations}
\begin{align}
&\mathcal{G}(s_1s_2s_3;s_1's_2's_3')\equiv\Big\{\Big[\big(s_1^2+s_2^2+s_3^2+s_1'^2+s_2'^2+s_3'^2\big)^2\notag\\
&-2\big(5s_1^2s_1'^2\!-\!s_1^2s_2'^2\!-\!s_1^2s_3'^2+5s_2^2s_2'^2\!-s_2^2s_1'^2\!-s_2^2s_3'^2+5s_3^2s_3'^2\notag\\
&-\!s_3^2s_1'^2\!\!-\!s_3^2s_2'^2\big)\Big]^2\!-36\,\big(2s_1^2s_2^2\!+\!2s_2^2s_3^2\!+\!2s_3^2s_1^2\!-\!s_1^4\!-\!s_2^4\!-\!s_3^4\big)\notag\\
&\times\big(2s_1'^2s_2'^2+2s_2'^2s_3'^2+2s_3'^2s_1'^2-s_1'^4-s_2'^4-s_3'^4\big)\Big\}^{-1/2}\label{eq:G2}
\end{align}
is the translationally and rotationally invariant Green function, satisfying
[$s_i$'s are defined in Eq.~\eqref{eq:s}]
\begin{equation}\label{eq:G1}
\sum_{i=1}^3\nabla_i^2\mathcal{G}(s_1s_2s_3;s_1's_2's_3')=-\frac{\pi}{2\sqrt{3}\,s_1's_2's_3'}\prod_{i=1}^3\delta(s_i-s_i').
\end{equation}
\end{subequations}
Because $\phi^{(3)}(\vect r_1\vect r_2\vect r_3)$ vanishes on $\mathcal{B}$,
the unknown function $F(R',c')$ in Eq.~\eqref{eq:phi3HS} satisfies an integral equation:
\begin{equation}\label{eq:IntegralEquationF}
\phi^{(3)}(\vect r_1\vect r_2\vect r_3)=0\text{~~~at~~}s_1=1,s_{2,3}=s_{\mp}(R,c),
\end{equation}
where $-1<c<1$ and $R>R_{\min}(c)$.

From Eqs.~\eqref{eq:phi3HS} and \eqref{eq:G1}, we get
\begin{equation}\label{eq:FRc_delta}
\big(\nabla_r^2+3\nabla_R^2/4\big)\phi^{(3)}(\vect r/2,-\vect r/2,\vect R)=F(R,c)\delta(r-1)
\end{equation}
at $\lvert\vect R\pm\vect r/2\rvert>1$, where $c=\hat{\vect R}\cdot\hat{\vect r}$.

From Eq.~\eqref{eq:FRc_delta}, it is clear that $F(R,-c)=F(R,c)$.

The 2-boson special functions satisfy Eqs.~\eqref{eq:2bodyspecial} at $r\ge1$ and vanish at $r\le1$, so
\begin{gather*}
a=1,~~r_s=2/3,~~r_s'=8/15,~~a_d=1/45,~~r_d=-150/7,\\
a_g=1/99225,~~~a_i=1/1404728325.
\end{gather*}
Applying $\big(\nabla_r^2+3\nabla_R^2/4\big)$ to Eq.~\eqref{eq:phi3kq_res_coordinate_final},
and comparing the result with Eq.~\eqref{eq:FRc_delta}, we get the asymptotic expansion of $F(R,c)$ at large $R$:
\begin{subequations}\label{eq:Fasymp}
\begin{equation}\label{eq:FasympRes}
F(R,c)=F^{(1)}_{a}(R,c)+QF^{(2)}_{a}(R,c)+O(R^{-8}),
\end{equation}
\begin{widetext}
\begin{multline}
F^{(1)}_{a}(R,c)=1-\frac{2}{R}+\frac{2w}{\pi R^2}-\frac{4w}{\pi R^3}
+\Big(\frac{24\sqrt{3}w}{\pi^2}\ln R+\frac{12}{\pi^2}-\frac{8}{\sqrt{3}\,\pi}-\frac{14}{9}\Big)R^{-4}
+\Big(\frac{96\sqrt{3}w}{\pi^2}\ln R+\frac{460}{9}+\frac{48}{\pi^2}-\frac{161}{\sqrt{3}\,\pi}\Big)R^{-5}\\
+\bigg[\frac{4(72+43\sqrt{3}\,\pi)w}{\pi^3}\ln R+\frac{10885}{54}+\frac{48\sqrt{3}}{\pi^3}+\frac{105}{\pi^2}
-\frac{5005}{8\sqrt{3}\,\pi}\bigg]R^{-6}
+\bigg[\frac{64(7\sqrt{3}\,\pi-9)w}{\pi^3}\ln R+\frac{61279}{135}-\frac{96\sqrt{3}}{\pi^3}+\frac{492}{\pi^2}\\-\frac{27733}{20\sqrt{3}\,\pi}\bigg]R^{-7}
+\bigg\{-\frac{5}{2R^3}+\Big(\frac{80}{3}-\frac{40\sqrt{3}}{\pi}\Big)R^{-4}-\frac{5w}{\pi R^5}+\Big(\frac{730}{\sqrt{3}\,\pi}-\frac{820}{9}
-\frac{360}{\pi^2}\Big)R^{-6}+\bigg[-\frac{120\sqrt{3}w}{\pi^2}\ln R\\
+\frac{15865}{4\sqrt{3}\,\pi}-\frac{4115}{9}-\frac{2220}{\pi^2}\bigg]R^{-7}\bigg\}P_2(c)
+\bigg[-\frac{9}{8R^5}+\Big(\frac{3648}{35}-\frac{1296\sqrt{3}}{7\pi}\Big)\frac{1}{R^6}-\frac{9w}{4\pi R^7}\bigg]P_4(c)
-\frac{13P_6(c)}{32R^7},
\end{multline}
\begin{equation}
F^{(2)}_{a}(R,c)=-\frac{6\sqrt{3}}{\pi R^4}-\frac{24\sqrt{3}}{\pi R^5}-\Big(\frac{72}{\pi^2}+\frac{43\sqrt{3}}{\pi}\Big)R^{-6}
+\Big[\frac{144}{\pi^2}-\frac{112\sqrt{3}}{\pi}+\frac{30\sqrt{3}P_2(c)}{\pi}\Big]R^{-7},
\end{equation}
\end{widetext}
\end{subequations}
where $Q$ is directly related to $D$:
\begin{align}\label{eq:DQrelation}
D&=48\pi^2Q-192\pi w(1-\gamma)-12\pi^2\notag\\
&=473.741~011~252~289Q-744.946~799~290~500.
\end{align}
It can be shown from Eqs.~\eqref{eq:IntegralEquationF} (at large $R$) and \eqref{eq:Fasymp} that
\begin{align}
Q&=\lim_{R\rightarrow\infty}\int_0^1\dif c'\int_{R_{\min}(c')}^{R}F(R',c')R'^2\dif R'\notag\\
&\quad\quad~\,-\big[R^3/3-R^2+2wR/\pi-(4w/\pi)\ln R\big].\label{eq:IntegralEquationQ}
\end{align}
Solving Eqs.~\eqref{eq:IntegralEquationF} and \eqref{eq:IntegralEquationQ} for $F(R,c)$ and $Q$ numerically,
[using $F^{(1)}_a(R',c')+QF^{(2)}_a(R',c')$ to approximate $F(R',c')$ for $R'$ greater than some sufficiently large value
in these two equations, and discretizing $F(R',c')$ for $R'$ less than this value], the author finds
$$Q=5.290~844\pm0.000~005.$$
Substituting this result into \eqref{eq:DQrelation}, we obtain Eq.~\eqref{eq:DHS}.

\section{Ground state of 3 bosons in a periodic cubic volume\label{sec:cubic}}
In this section we return to the general interactions considered in Sec.~\ref{sec:asymp}.
The 3 bosons are now placed in a large periodic cubic volume of side $L$, and
\begin{equation}
\epsilon\equiv1/L
\end{equation}
is the small parameter.

\subsection{Ground state wave function and energy}
Let $A_{\vect n_1\vect n_2\vect n_3}$ ($\sum_{i=1}^3\vect n_i\equiv0$) be proportional to the probability amplitude that
the bosons have momenta $2\pi\epsilon\vect n_i$ ($i=1,2,3$). We shall call $\vect n_i$ an \emph{integral vector},
since its Cartesian components (along the sides of the cubic volume) are integers.

Schr\"{o}dinger equation ($\hbar=m_\text{boson}=1$)
\begin{align}\label{eq:A}
&\big[2\pi^2\epsilon^2(n_1^2+n_2^2+n_3^2)-E\big]A_{\vect n_1\vect n_2\vect n_3}\notag\\
&+\frac{1}{6L^6}U_{2\pi\epsilon\vect n_1,2\pi\epsilon\vect n_2,2\pi\epsilon\vect n_3,
2\pi\epsilon\vect n_1',2\pi\epsilon\vect n_2',2\pi\epsilon\vect n_3'}A_{\vect n_1'\vect n_2'\vect n_3'}\notag\\
&+\frac{1}{2L^3}U_{2\pi\epsilon\vect n_1,2\pi\epsilon\vect n_2,2\pi\epsilon\vect n',2\pi\epsilon\vect n''}A_{\vect n'\vect n''\vect n_3}
+\cdots=0,
\end{align}
where summation over dumb momenta is implicit,
and ``$\cdots$" stands for two other similar pairwise interaction terms, is rewritten as
\begin{align}
&\big[(k_1^2+k_2^2+k_3^2)/2-E\big]\psi_{\vect k_1\vect k_2\vect k_3}\notag\\
&+\frac{1}{6}I_\epsilon(\vect k_1)I_\epsilon(\vect k_2)\int_{\vect k_1'\vect k_2'}
U_{\vect k_1\vect k_2\vect k_3\vect k_1'\vect k_2'\vect k_3'}\psi_{\vect k_1'\vect k_2'\vect k_3'}\notag\\
&+\frac{1}{2}I_\epsilon(\vect k_1)\int_{\vect k'}U_{\vect k_1\vect k_2\vect k'\vect k''}\psi_{\vect k'\vect k''\vect k_3}
+\cdots=0,\label{eq:psi_Schrodinger}
\end{align}
\begin{equation}\label{eq:psi}
\psi_{\vect k_1\vect k_2\vect k_3}\equiv\sum_{\vect n_1\vect n_2}A_{\vect n_1\vect n_2\vect n_3}
\prod_{i=1}^2(2\pi\epsilon)^3\delta(\vect k_i-2\pi\epsilon\vect n_i),
\end{equation}
where $\sum_{i=1}^3\vect k_i\equiv0$,
$\int_{\vect k'}\equiv\int\frac{\dif^3k'}{(2\pi)^3}$, $\int_{\vect k_1'\vect k_2'}\equiv\int\int\frac{\dif^3k_1'}{(2\pi)^3}\frac{\dif^3k_2'}{(2\pi)^3}$,
and
\begin{align}
I_\epsilon(\vect k)&\equiv\sum_{\vect n}(2\pi\epsilon)^3\delta(\vect k-2\pi\epsilon\vect n).\\
J_\epsilon(\vect k)&\equiv\sum_{\vect n\neq0}(2\pi\epsilon)^3\delta(\vect k-2\pi\epsilon\vect n).
\end{align}

In the following, $\vect k$'s will be taken as independent of $\epsilon$ but $\vect q$'s be small momenta such that $\vect q/\epsilon$'s are independent
of $\epsilon$. Also, $I_\epsilon(\vect k)$ and $J_\epsilon(\vect k)$ will be simply written as
$I(\vect k)$ and $J(\vect k)$, respectively.

Let
\begin{equation}\label{eq:A000}
A_{0,0,0}\equiv L^6,
\end{equation}
so $\psi_{\vect q_1\vect q_2\vect q_3}=\prod_{i=1}^2(2\pi)^3\delta(\vect q_i)+o(\epsilon^{-6})$.

The following expansions will be found:
\begin{subequations}
\begin{align}
\psi_{\vect k_1\vect k_2\vect k_3}&=\sum_{s\ge0}\CR^{(s)}_{\vect k_1\vect k_2\vect k_3},\label{eq:CRdef}\\
\psi_{\vect q,-\vect q/2+\vect k,-\vect q/2-\vect k}&=\sum_{s\ge-3}\CS^{(s)\vect q}_\vect k,\label{eq:CSdef}\\
\psi_{\vect q_1\vect q_2\vect q_3}&=\sum_{s\ge-6}\CT^{(s)}_{\vect q_1\vect q_2\vect q_3},\label{eq:CTdef}\\
E&=\sum_{s\ge3}E^{(s)},\label{eq:Edef}
\end{align}
\end{subequations}
where $\CR^{(s)}_{\vect k_1\vect k_2\vect k_3}$, $\CS^{(s)\vect q}_\vect k$,
$\CT^{(s)}_{\vect q_1\vect q_2\vect q_3}$, and $E^{(s)}$ scale with $\epsilon$ like $\epsilon^s$ (\emph{not} excluding $\epsilon^s\ln\epsilon$).
Equation~\eqref{eq:CRdef} is understood in the following sense: we expand the Fourier transform of $\psi_{\vect k_1\vect k_2\vect k_3}$
within a large but \emph{fixed} spatial region in powers of $\epsilon$, and then transform the result back to the $\vect k_i$-space term by term,
to obtain $\sum_s\CR^{(s)}_{\vect k_1\vect k_2\vect k_3}$. Equation~\eqref{eq:CSdef} is similar: we expand the partial Fourier transform,
$\int_{\vect k}\psi_{\vect q,-\vect q/2+\vect k,-\vect q/2-\vect k}\exp(\I\vect k\cdot\vect r)$, within a large but fixed
region of the $\vect r$-space, for fixed $\vect q/\epsilon$, in powers of $\epsilon$, and then transform
the result back to the $\vect k$-space term by term, to obtain $\sum_s\CS^{(s)\vect q}_\vect k$.

Obviously
\begin{equation}\label{eq:CT-6}
\CT^{(-6)}_{\vect q_1\vect q_2\vect q_3}=(2\pi)^3\delta(\vect q_1)(2\pi)^3\delta(\vect q_2).
\end{equation}
Therefore $\CT^{(-6)}_{\vect k_1\vect k_2\vect k_3}=(2\pi)^3\delta(\vect k_1)(2\pi)^3\delta(\vect k_2)$
and $\CT^{(-6)}_{\vect q,-\vect q/2+\vect k,-\vect q/2-\vect k}=(2\pi)^3\delta(\vect q)(2\pi)^3\delta(\vect k)$,
indicating that the minimum values of $s$ for $\CR^{(s)}_{\vect k_1\vect k_2\vect k_3}$ and $\CS^{(s)\vect q}_\vect k$ are 0 and $-3$, respectively.
Since the energy is dominated by pairwise mean-field interactions, $E\sim\epsilon^3$.

The Fourier transform of $I(\vect k)$ becomes $\delta(\vect r)$ within any \emph{fixed} spatial region,
when $\epsilon$ is sufficiently small. Using the prescription in Appendix~\ref{sec:Z-delta}, we thus get:
$$
I(\vect k)=1+O(\epsilon^{s_c}) \text{~~~for \emph{any} $s_c$}.
$$
Using this result, and also noting that the interactions are finite-ranged, we get
three specialized forms of Eq.~\eqref{eq:psi_Schrodinger}
(accurate to any finite order in $\epsilon$):
\begin{subequations}\label{eq:CRCSCT}
\begin{align}
&\frac{1}{2}(k_1^2+k_2^2+k_3^2)\CR_{\vect k_1\vect k_2\vect k_3}
+\frac{1}{6}\int_{\vect k_1'\vect k_2'}U_{\vect k_1\vect k_2\vect k_3\vect k_1'\vect k_2'\vect k_3'}
\CR_{\vect k_1'\vect k_2'\vect k_3'}\notag\\
&+\frac{1}{2}\int_{\vect k'}U_{\vect k_1\vect k_2\vect k'\vect k''}\CR_{\vect k_3\vect k'\vect k''}
+\cdots=E\CR_{\vect k_1\vect k_2\vect k_3},\label{eq:CR}
\end{align}
\begin{equation}\label{eq:CS}
(H\CS^\vect q)_{\vect k}+\big({3}q^2/{4}-E\big)\CS^\vect q_{\vect k}+I(\vect q)\mathcal{W}^\vect q_\vect k=0,
\end{equation}
\begin{align}
&\Big[\frac{1}{2}(q_1^2+q_2^2+q_3^2)-E\Big]\CT_{\vect q_1\vect q_2\vect q_3}
+\frac{1}{2}\sum_{i=1}^3I(\vect q_{i'}^{})\!\int_{\vect k'}U_{\vect p_i\vect k'}\CS^{\vect q_i}_{\vect k'}\notag\\
&+\frac{1}{6}I(\vect q_1)I(\vect q_2)\int_{\vect k_1'\vect k_2'}
U_{\vect q_1\vect q_2\vect q_3\vect k_1'\vect k_2'\vect k_3'}\CR_{\vect k_1'\vect k_2'\vect k_3'}=0,\label{eq:CT}
\end{align}
\end{subequations}
\begin{align}
&\mathcal{W}^\vect q_\vect k\equiv\frac{1}{6}\int_{\vect k_1'\vect k_2'}
U_{-\vect q/2+\vect k,-\vect q/2-\vect k,\vect q,\vect k_1'\vect k_2'\vect k_3'}\CR_{\vect k_1'\vect k_2'\vect k_3'}\notag\\
&+\!\Big[\frac{1}{2}\!\int_{\vect k'}\!\!U_{-\vect q/2+\vect k,\vect q,\vect k'\vect k''}\CR_{-\vect q/2-\vect k,\vect k'\vect k''}
\!+\!(\vect q\leftrightarrow-\vect q)\Big],
\end{align}
where $\CR$, $\CS$, and $\CT$ are the sums of $\CR^{(s)}$, $\CS^{(s)}$, and $\CT^{(s)}$, respectively,
and in Eq.~\eqref{eq:CT}, $1'=2$, $2'=3$, $3'=1$, and $\vect p_i$ is defined in Eq.~\eqref{eq:p}.

The following \textit{statement} is now true at $s_1=-6$:

\textbf{\textit{Statement} $\mathbf{s_1}$}:
The functions $\CT^{(s)}_{\vect q_1\vect q_2\vect q_3}$ (for $s\le s_1$),
$\CS^{(s)\vect q}_\vect k$ (for $s\le s_1+2$), $\CR^{(s)}_{\vect k_1\vect k_2\vect k_3}$ (for $s\le s_1+5$),
and $E^{(s)}$ (for $s\le s_1+8$), have all been formally determined.

Now do the following expansions for $-6\le s\le s_1$:
\begin{subequations}
\begin{align}
\CT^{(s)}_{\vect q,-\vect q/2+\vect k,-\vect q/2-\vect k}&=\sum_nt'^{(n,s-n)}_{\vect q,\vect k},\\
\CT^{(s)}_{\vect k_1\vect k_2\vect k_3}&=\sum_nt''^{(n,s-n)}_{\vect k_1\vect k_2\vect k_3},
\end{align}
\end{subequations}
where $\vect q/\epsilon$'s and $\vect k$'s are indepedent of $\epsilon$,
$\vect k_i\sim k$, and $t'^{(n,s-n)}_{\vect q,\vect k}\sim\epsilon^nk^{s-n}$,
$t''^{(n,s-n)}_{\vect k_1\vect k_2\vect k_3}\sim\epsilon^nk^{s-n}$.
For the same reason as Eq.~\eqref{eq:S_eq1},
we obtain the following asymptotic expansions at small $\vect k$'s:
\begin{subequations}
\begin{align}
\CS^{(s_1+3)\vect q}_\vect k&=\sum_{m=-s_1-9}^{-3}t'^{(s_1+3,m)}_{\vect q,\vect k}+O(\epsilon^{s_1+3}k^{-2}),\label{eq:CSsmallk}\\
\CR^{(s_1+6)}_{\vect k_1\vect k_2\vect k_3}&=\sum_{m=-s_1-12}^{-6}t''^{(s_1+6,m)}_{\vect k_1\vect k_2\vect k_3}+O(\epsilon^{s_1+6}k^{-5}).
\label{eq:CRsmallk}
\end{align}
\end{subequations}

Extracting all the terms that scale like $\epsilon^{s_1+6}$ from Eq.~\eqref{eq:CR}, we get
\begin{align}
&\frac{1}{2}(k_1^2+k_2^2+k_3^2)\CR^{(s_1+6)}_{\vect k_1\vect k_2\vect k_3}
+\frac{1}{6}\int_{\vect k_1'\vect k_2'}U_{\vect k_1\vect k_2\vect k_3\vect k_1'\vect k_2'\vect k_3'}
\CR^{(s_1+6)}_{\vect k_1'\vect k_2'\vect k_3'}\notag\\
&+\frac{1}{2}\int_{\vect k'}U_{\vect k_1\vect k_2\vect k'\vect k''}\CR^{(s_1+6)}_{\vect k_3\vect k'\vect k''}
+\cdots=\!\!\!\sum_{3\le n\le s_1+6}\!\!\!\!E^{(n)}\CR^{(s_1+6-n)}_{\vect k_1\vect k_2\vect k_3},\label{eq:CRs1p6}
\end{align}
where the right hand side is already known.
This equation and Eq.~\eqref{eq:CRsmallk} [which expands $\CR^{(s_1+6)}_{\vect k_1\vect k_2\vect k_3}$
to the order $k^{-6}$ at small $\vect k$'s] are sufficient to determine
$\CR^{(s_1+6)}_{\vect k_1\vect k_2\vect k_3}$, due to the 3-body version of the uniqueness theorem
[analogous to Eq.~\eqref{eq:unique2}].

The above information for $\CS^{\vect q}_\vect k$ and $\CR_{\vect k_1\vect k_2\vect k_3}$
is sufficient to determine each term in Eq.~\eqref{eq:CS}, except the first term, to the order $\epsilon^{s_1+3}$
[note that $I(\vect q)\sim\epsilon^0$], so $(HS^{(s_1+3)\vect q})_\vect k$ is known; taking into account
Eq.~\eqref{eq:CSsmallk}, one can determine $\CS^{(s_1+3)\vect q}_\vect k$.

The sum of all the interaction terms in Eq.~\eqref{eq:CT} can now be determined to the order $\epsilon^{s_1+3}$,
with a result $c'_{\vect q_1\vect q_2\vect q_3}+c_0\prod_{i=1}^2(2\pi\epsilon)^3\delta(\vect q_i)$,
where $c'_{\vect q_1\vect q_2\vect q_3}$ vanishes at $\vect q_1=\vect q_2=0$.
Both $c'_{\vect q_1\vect q_2\vect q_3}$ and $c_0$ are known up to the order $\epsilon^{s_1+3}$. 

Because of Eq.~\eqref{eq:A000},
$\CT'_{\vect q_1\vect q_2\vect q_3}\equiv\CT_{\vect q_1\vect q_2\vect q_3}-\prod_{i=1}^2(2\pi)^3\delta(\vect q_i)$
vanishes at $\vect q_1=\vect q_2=0$. Solving Eq.~\eqref{eq:CT}, we get $E=c_0\epsilon^6$ (so now $E$ is known to the order $\epsilon^{s_1+9}$)
and
\begin{equation}\label{eq:CTprime}
\CT'_{\vect q_1\vect q_2\vect q_3}=G_{q_1q_2q_3}(E\CT'_{\vect q_1\vect q_2\vect q_3}-c'_{\vect q_1\vect q_2\vect q_3}),
\end{equation}
where $G_{q_1q_2q_3}\equiv2/(q_1^2+q_2^2+q_3^2)\sim\epsilon^{-2}$.
Because $E\sim\epsilon^3$, $\CT'_{\vect q_1\vect q_2\vect q_3}\sim\epsilon^x$, $x>-6$,
and $E$ and $\CT'_{\vect q_1\vect q_2\vect q_3}$ are known to the orders $\epsilon^{s_1+9}$ and $\epsilon^{s_1}$, respectively,
we know $E\CT'_{\vect q_1\vect q_2\vect q_3}$ to the order $\epsilon^{s_1+3}$. 
So from Eq.~\eqref{eq:CTprime}, we can now
determine $\CT'_{\vect q_1\vect q_2\vect q_3}$ (and thus $\CT_{\vect q_1\vect q_2\vect q_3}$) to the order $\epsilon^{s_1+1}$.

The truth of \textit{Statement} $(s_1+1)$ is now established.

Repeating the above routine, one can formally determine $A_{\vect n_1\vect n_2\vect n_3}$ and $E$ to any orders
in $\epsilon$. The following are the step-by-step results of this program.

\textbf{\textit{Step 1a.}}
$\CR^{(0)}$, $\CR^{(1)}$, and $\CR^{(2)}$ satisfy the \emph{zero}-energy Schr\"{o}dinger equation [because of Eq.~\eqref{eq:CRs1p6}].
They will all be determined in this paper. From Eq.~\eqref{eq:CT-6} we get $t''^{(0,-6)}_{\vect k_1\vect k_2\vect k_3}
=\prod_{i=1}^2(2\pi)^3\delta(\vect k_i)$. So
\begin{equation}
\CR^{(0)}_{\vect k_1\vect k_2\vect k_3}=\phi^{(3)}_{\vect k_1\vect k_2\vect k_3}.
\end{equation}

\textbf{\textit{Step 1b.}}
$t'^{(-3,-3)}_{\vect q,\vect k}=(2\pi)^3\delta(\vect q)(2\pi)^3\delta(\vect k)$; $(H\CS^{(-3)\vect q})_\vect k=0$. So
\begin{equation}
\CS^{(-3)\vect q}_\vect k=(2\pi)^3\delta(\vect q)\phi_\vect k.
\end{equation}

\textbf{\textit{Step 2.}}
With the help of Eqs.~\eqref{eq:intUphi}, and noting that $\sum_{i=1}^3\vect q_i\equiv0$, we get
\begin{align}
E&=12\pi a\epsilon^3+o(\epsilon^3),\\
\CT^{(-5)}_{\vect q_1\vect q_2\vect q_3}&=\sum_{i=1}^3-(4\pi a/p_i^2)J(\vect p_i)(2\pi)^3\delta(\vect q_i).
\end{align}
Using the method of Appendix~\ref{sec:Z-delta}, we get
\begin{subequations}\label{eq:CT-5expand}
\begin{align}
&\CT^{(-5)}_{-\vect q/2+\vect k,-\vect q/2-\vect k,\vect q}=J(\vect q)\big[-8\pi a/q^2-\pi a(\hat{\vect q}\cdot\nabla_k)^2\notag\\
&+O(\epsilon^2)\big](2\pi)^3\delta(\vect k)
+(2\pi)^3\delta(\vect q)\big[-4\pi a/k^2\notag\\&-\alpha_1a\epsilon(2\pi)^3\delta(\vect k)/\pi
+2\pi a\epsilon^3\nabla_k^2(2\pi)^3\delta(\vect k)/3+O(\epsilon^5)\big],
\end{align}
\begin{align}
&\CT^{(-5)}_{\vect k_1\vect k_2\vect k_3}=\big[\sum_{i=1}^3-(4\pi a/l_i^2)(2\pi)^3\delta(\vect k_i)\big]\notag\\
&\quad~~-(3\alpha_1a\epsilon/\pi)(2\pi)^3\delta(\vect k_1)(2\pi)^3\delta(\vect k_2)+O(\epsilon^3),
\end{align}
\end{subequations}
\begin{equation}\label{eq:l}
\vect l_1\equiv(\vect k_2-\vect k_3)/2,\text{~~~and similarly for $\vect l_{2}, \vect l_3$}.
\end{equation}
\begin{subequations}
\begin{gather}
\alpha_s\equiv\lim_{\eta\rightarrow0^+}\sum_{\vect n\ne0}\frac{\e^{-\eta n}}{\lvert\vect n\rvert^{2s}}
-\int\dif^3n\frac{\e^{-\eta n}}{\lvert\vect n\rvert^{2s}}~~~(s<3/2),\\
\alpha_{1.5}^{}\equiv\lim_{N\rightarrow\infty}[(\sum_{\vect n\ne0; n<N}\lvert\vect n\rvert^{-3})-4\pi\ln N],\\
\alpha_s\equiv\sum_{\vect n\ne0}\lvert\vect n\rvert^{-2s}~~~(s>3/2).
\end{gather}
\end{subequations}
The above lattice sums are evaluated in Appendix~\ref{sec:constants}.
One can easily extract $t'^{(j,-5-j)}_{\vect q,\vect k}$ and $t''^{(j,-5-j)}_{\vect k_1\vect k_2\vect k_3}$ from
Eqs.~\eqref{eq:CT-5expand}.

\textbf{\textit{Step 3a.}}
From $t''^{(1,-6)}_{\vect k_1\vect k_2\vect k_3}$ and Eq.~\eqref{eq:CRs1p6}, we get
\begin{equation}
\CR^{(1)}_{\vect k_1\vect k_2\vect k_3}=-(3\alpha_1a\epsilon/\pi)\phi^{(3)}_{\vect k_1\vect k_2\vect k_3}.
\end{equation}

\textbf{\textit{Step 3b.}}
From $(H\CS^{(-2)\vect q})_\vect k=0$, and $t'^{(-2,-3)}_{\vect q,\vect k}$, we get
\begin{equation}
\CS^{(-2)\vect q}_\vect k=\big[-(\alpha_1a\epsilon/\pi)(2\pi)^3\delta(\vect q)-8\pi aJ(\vect q)/q^2\big]\phi_\vect k.
\end{equation}

\textbf{\textit{Step 4.}}
This is similar to \textit{Step 2}.
\begin{equation}
E=12\pi a\epsilon^3(1-\alpha_1a\epsilon/\pi)+O(\epsilon^{5}),
\end{equation}
\begin{align}
&\CT^{(-4)}_{\vect q_1\vect q_2\vect q_3}=J_{\vect q_1\vect q_2\vect q_3}G_{n_1n_2n_3}\sum_{i=1}^32a^2/\pi^2\epsilon^4n_i^2\notag\\
&~+(a^2/\pi^2\epsilon)\sum_{i=1}^3(\alpha_1^{}m_i^{-2}+m_i^{-4})J(\vect p_i)(2\pi)^3\delta(\vect q_i),\label{eq:CT-4}
\end{align}
\begin{equation}
J_{\vect q_1\vect q_2\vect q_3}\equiv\sum_{\vect n'_{1,2}\ne0,\vect n'_1+\vect n'_2\ne0}\prod_{i=1}^2
(2\pi\epsilon)^3\delta(\vect q_i-2\pi\epsilon\vect n'_i).
\end{equation}
The three subscripts of $J_{\vect q_1\vect q_2\vect q_3}$ are always subject to the constraint $\sum_{i=1}^3\vect q_i\equiv0$.
In Eq.~\eqref{eq:CT-4} and in the following, $G_{n_1n_2n_3}\equiv2/(n_1^2+n_2^2+n_3^2)$, and
\begin{equation}
\vect n_i\equiv\vect q_i/2\pi\epsilon,~~~\vect n\equiv\vect q/2\pi\epsilon,~~~\vect m_i\equiv\vect p_i/2\pi\epsilon.
\end{equation}

$\CT^{(-4)}_{\vect q,-\vect q/2+\vect k,-\vect q/2-\vect k}=\sum_{j=-2}^{\infty}t'^{(j,-4-j)}_{\vect q,\vect k}$,
where
\begin{subequations}
\begin{align}
&t'^{(-1,-3)}_{\vect q,\vect k}=(a^2/\pi^2\epsilon)\big\{(\alpha_1^2+\alpha_2^{})(2\pi\epsilon)^3\delta(\vect q)(2\pi)^3\delta(\vect k)\notag\\
&\quad~+J(\vect q)\big[2\rho_{A1}^{}(\vect n)+2\alpha_1^{}/n^2\!-\!6/n^4\big](2\pi)^3\delta(\vect k)\big\},\\
&t'^{(0,-4)}_{\vect q,\vect k}\!=\big[16\pi^2a^2(2\pi\epsilon)^3\delta(\vect q)+40\pi^2a^2J(\vect q)\big]Z/k^4,\\
&t'^{(1,-5)}_{\vect q,\vect k}=X^{(-4)}_{\vect q,\vect k}J(\vect q).
\end{align}
\end{subequations}
The actual formulas for $X^{(s)}_{\vect q,\vect k}$ ($s=-4,-3,-2$; see \textit{Steps 6} and \textit{8} below)
are not needed in this paper. For $s\ge1$,
\begin{subequations}
\begin{align}
\rho_{As}^{}(\vect n)&\equiv2\theta_{As}(\vect n)+W_A(\vect n)/n^{2s},\label{eq:rhoA1_def}\\
\theta_{As}(\vect n)&\equiv\sum_{\vect m\ne0}\big[(m^2+\vect m\cdot\vect n+n^2)m^{2s}\big]^{-1},\\
W_A(\vect n)&\equiv\lim_{N\rightarrow\infty}\sum_{\lvert\vect m\rvert<N}(m^2+\vect m\cdot\vect n+n^2)^{-1}-4\pi N.
\end{align}
\end{subequations}
\begin{align}
&\CT^{(-4)}_{\vect k_1\vect k_2\vect k_3}\!=G_{k_1k_2k_3}\sum_{i=1}^3{32\pi^2a^2}/{k_i^2}\notag\\
&\quad\quad\quad\quad+12\alpha_1a^2\epsilon\sum_{i=1}^3{(2\pi)^3\delta(\vect k_i)}/{l_i^2}\notag\\
&+3\pi^{-2}(2\beta_{1A}+\alpha_1^2-3\alpha_2^{})a^2\epsilon^2(2\pi)^6\delta(\vect k_1)\delta(\vect k_2)+O(\epsilon^3),
\end{align}
from which one can easily extract $t''^{(j,-4-j)}_{\vect k_1\vect k_2\vect k_3}$. Here
\begin{equation}
\beta_{1A}\equiv\lim_{\eta\rightarrow0^+}\sum_{\vect n\ne0}\e^{-\eta n}W_A(\vect n)/n^2+4\sqrt{3}\,\pi^3/\eta^2
\end{equation}
is evaluated in Eq.~\eqref{eq:beta1A_numerical}.

\textbf{\textit{Step 5a.}}
From $t''^{(2,-6)}_{\vect k_1\vect k_2\vect k_3}$ and Eq.~\eqref{eq:CRs1p6}, we get
\begin{equation}
\CR^{(2)}_{\vect k_1\vect k_2\vect k_3}=3\pi^{-2}(2\beta_{1A}+\alpha_1^2-3\alpha_2^{})a^2\epsilon^2\phi^{(3)}_{\vect k_1\vect k_2\vect k_3}.
\end{equation}

\textbf{\textit{Step 5b.}}
From $(H\CS^{(-1)\vect q})_\vect k=0$, and $t'^{(-1,-3)}_{\vect q,\vect k}$, we get
\begin{align}
\CS^{(-1)\vect q}_\vect k&=(a^2/\pi^2\epsilon)\big\{(\alpha_1^2+\alpha_2)(2\pi\epsilon)^3\delta(\vect q)\notag\\
&\quad+\big[2\rho_{A1}^{}(\vect n)+2\alpha_1/n^2-6/n^4\big]J(\vect q)\big\}\phi_{\vect k}.
\end{align}

\textbf{\textit{Step 6.}}
This is similar to \textit{Steps 2} and \textit{4}.
\begin{equation}
E=12\pi a\epsilon^3\big[1-\alpha_1a\epsilon/\pi+(\alpha_1^2+\alpha_2)a^2\epsilon^2/\pi^2\big]+O(\epsilon^{6}),
\end{equation}
\begin{align}
&\CT^{(-3)}_{\vect q_1\vect q_2\vect q_3}=\sum_{i=1}^3(6a^3/\pi^3\epsilon^3)J_{\vect q_1\vect q_2\vect q_3}G_{n_1n_2n_3}^2n_i^{-2}\notag\\
&-(2a^3/\pi^3\epsilon^3)J_{\vect q_1\vect q_2\vect q_3}G_{n_1n_2n_3}
\big[\rho_{A1}^{}(\vect n_i)+\alpha_1/n_i^2-3/n_i^4\big]\notag\\
&+\big\{\big[\!-\!4\rho_{A1}^{}(\vect m_i)/m_i^2\!-\!(\alpha_1^2\!+\!\alpha_2)/m_i^2
+2\alpha_1/m_i^4\!+15/m_i^6\big]\notag\\
&\quad\quad\quad\quad\quad\quad\quad\times a^3/\pi^3+u_0\big\}J(\vect p_i)(2\pi)^3\delta(\vect q_i).
\end{align}
\begin{subequations}

In preparation for the subsequent steps, we derive
\begin{align}
&t'^{(0,-3)}_{\vect q,\vect k}=\big\{-32\pi^2wa^3Z_{1/\lvert a\rvert}(k)/k^3\notag\\
&\quad+\big[2u_0+16wa^3\ln(2\pi\epsilon\lvert a\rvert)\big](2\pi)^3\delta(\vect k)\notag+(2a^3/\pi^3)\\
&\quad\times\big[-\rho_{AA1}^{}(\vect n)-\alpha_1^{}\rho_{A1}^{}(\vect n)+3\rho_{A2}^{}(\vect n)+3\rho_{B1}^{}(\vect n)\notag\\
&\quad-(\alpha_1^2+\alpha_2)/n^2+6\alpha_1/n^4-9/n^6\big](2\pi)^3\delta(\vect k)
\big\}J(\vect q)\notag\\
&+\big\{-32\pi^2wa^3Z_{1/\lvert a\rvert}(k)/k^3+\big[16wa^3\ln(2\pi\epsilon\lvert a\rvert)-u_0\notag\\
&\quad\,\,\,\,\,+(15\alpha_3+\alpha_1\alpha_2-\alpha_1^3-4\alpha_{1A1}^{})a^3/\pi^3\big]
(2\pi)^3\delta(\vect k)\big\}\notag\\
&\quad\quad\quad\quad\times(2\pi\epsilon)^3\delta(\vect q),\label{eq:t'0m3}
\end{align}
\begin{equation}
t'^{(1,-4)}_{\vect q,\vect k}\!\!\!=X^{(-3)}_{\vect q,\vect k}J(\vect q)-(96\pi\alpha_1a^3\epsilon Z/k^4)(2\pi\epsilon)^3\delta(\vect q),
\end{equation}
\end{subequations}
where $\rho_{B1}^{}(\vect n)$, $\rho_{AA1}^{}(\vect n)$, and $\alpha_{1A1}^{}$ are defined as follows.

For $s\ge0$,
\begin{subequations}
\begin{align}
\rho_{Bs}^{}(\vect n)&\equiv2\theta_{Bs}^{}(\vect n)+W_B^{}(\vect n)n^{-2s},\label{eq:rhoB1_def}\\
\theta_{Bs}^{}(\vect n)&\equiv\sum_{\vect m\ne0}\big[(m^2+\vect m\cdot\vect n+n^2)^2m^{2s}\big]^{-1},\\
W_B^{}(\vect n)&\equiv\sum_{\text{all }\vect m}(m^2+\vect m\cdot\vect n+n^2)^{-2}.
\end{align}
\end{subequations}
\begin{align}\label{eq:rhoAA1_def}
\rho_{AA1}^{}(\vect n)&\equiv\lim_{N\rightarrow\infty}\big[2\!\!\sum_{\vect m\ne0; m<N}(m^2+\vect m\cdot\vect n+n^2)^{-1}\rho_{A1}(\vect m)\notag\\
&\quad\quad\quad-8\pi^3w\ln N\big]+W_A(\vect n)\rho_{A1}(\vect n).
\end{align}
It can be shown that at large $\vect n$
\begin{align*}
\rho_{A1}^{}(\vect n)&=\pi^2w/n+2\alpha_1/n^2+O(n^{-4}),\\
\rho_{AA1}^{}(\vect n)&=-8\pi^3w\ln n-\pi^3w(7\pi/\sqrt{3}-8)+2\pi^2w\alpha_1/n\\
&\quad+O(n^{-2}).
\end{align*}
We define some 2-fold lattice sums:
\begin{subequations}
\begin{align}
\alpha_{1A1}^{}&\equiv\!\lim_{N\rightarrow\infty}\big\{[\!\!\sum_{\vect n\ne0;n<N}\!n^{-2}\rho_{A1}^{}(\vect n)]\!-\!4\pi^3w\ln N\big\},\\
\alpha_{sAs'}^{}&\equiv\sum_{\vect n\ne0}n^{-2s}\rho_{As'}^{}(\vect n)~~~(s+s'\ge3),\\
\alpha_{sBs'}^{}&\equiv\sum_{\vect n\ne0}n^{-2s}\rho_{Bs'}^{}(\vect n)~~~(s+s'\ge2).
\end{align}
\end{subequations}
$\alpha_{1A1}^{}$, $\alpha_{1A2}$, $\alpha_{2A1}$, and $\alpha_{1B1}$ are evaluated in Appendix~\ref{sec:constants}.

\textbf{\textit{Step 7.}}
From
\begin{align*}
&(H\CS^{(0)\vect q})_\vect k+(-6\pi a\phi_\vect k+W^{(0)}_\vect k)J(\vect q)\\
&\quad\quad+(-12\pi a\phi_\vect k+W^{(0)}_\vect k)(2\pi\epsilon)^3\delta(\vect q)=0~~~(\text{all }\vect k),\\
&\CS^{(0)\vect q}_\vect k=\sum_{s=-5}^{-3}t'^{(0,s)}_{\vect q,\vect k}+O(\epsilon^0k^{-2})~~~(\text{small }\vect k),
\end{align*}
where $W^{(0)}_\vect k$ is defined in Eq.~\eqref{eq:WTaylor}, we get
\begin{align}
&\CS^{(0)\vect q}_\vect k=\big\{\big[16wa^3\ln(\epsilon\lvert a\rvert)
+2a^3\widetilde{\rho}_{AA1}^{}(\vect n)/\pi^3\big]\phi_\vect k+d_\vect k\notag\\
&+10\pi a\phi^{(d)}_{\hat{\vect q}\vect k}
\big\}J(\vect q)+\big\{\big[16wa^3\ln(\epsilon\lvert a\rvert)+C_0a^3+3\pi a^2r_s\notag\\
&\quad\quad\quad\quad\quad-3u_0\big]\phi_\vect k+d_\vect k+6\pi af_\vect k\big\}(2\pi\epsilon)^3\delta(\vect q),
\end{align}
where $d_\vect k$ is defined in Eqs.~\eqref{eq:d},
\begin{align}
C_0&\equiv16w\ln(2\pi)+(15\alpha_3+\alpha_1\alpha_2-\alpha_1^3-4\alpha_{1A1}^{})/\pi^3\notag\\
&\quad-\big(14\pi/\sqrt{3}-16\big)w\notag\\
&=95.852~723~604~821~230~29,
\end{align}
and
\begin{align}
\widetilde{\rho}_{AA1}^{}(\vect n)\equiv&-\rho_{AA1}^{}(\vect n)-\pi^3w\big(7\pi/\sqrt{3}-8\big)+8\pi^3w\ln(2\pi)\notag\\
&-\alpha_1^{}\rho_{A1}^{}(\vect n)+3\rho_{A2}^{}(\vect n)+3\rho_{B1}^{}(\vect n)\notag\\
&-(\alpha_1^2+\alpha_2)/n^2+6\alpha_1/n^4-9/n^6.\label{eq:rhotAA1_def}
\end{align}
See Appendix~\ref{sec:constants} for the evaluation of $C_0$ (and $C_1$ in \textit{Step 9.}).

\textbf{\textit{Step 8.}}
This is similar to \textit{Steps 2}, \textit{4}, and \textit{6}. With the help of Eqs.~\eqref{eq:intUphi} and \eqref{eq:D_complicated}, we get
\begin{align}
&E=12\pi a\epsilon^3\big\{1-\alpha_1a\epsilon/\pi+(\alpha_1^2+\alpha_2)a^2\epsilon^2/\pi^2\notag\\
&\!+\!\!\big[16wa^3\ln(\epsilon\lvert a\rvert)+C_0a^3+3\pi a^2r_s\big]\epsilon^3\big\}\!+\!D\epsilon^6\!+\!O(\epsilon^7),\label{eq:E_L-6}
\end{align}
\begin{subequations}
\begin{align}
&\CT^{(-2)}_{\vect q_1\vect q_2\vect q_3}=\sum_{i=1}^3\big[J(\vect p_i)(2\pi)^3\delta(\vect q_i)\CT^{(-2\delta)}(\vect m_i)\notag\\
&\quad\quad\quad\quad~+J_{\vect q_1\vect q_2\vect q_3}\sum\nolimits_{s=0}^3G_{\vect n_1\vect n_2\vect n_3}^s\CT^{(-2)}_s(\vect n_i)\big],\\
&\CT^{(-2\delta)}(\vect m_i)\equiv-\alpha_1au_0\epsilon/\pi-au_0\epsilon/\pi m_i^2-D\epsilon/4\pi^2m_i^2\notag\\
&-3a^3r_s\epsilon/m_i^2+\big\{-\big[4\widetilde{\rho}_{AA1}^{}(\vect m_i)+48\pi^3w\ln(\epsilon\lvert a\rvert)\notag\\
&\quad\quad+\pi^3C_0\big]/m_i^2-\big[12\rho_{A1}^{}(\vect m_i)+9\alpha_1^2+6\alpha_2\big]/m_i^4\notag\\
&\quad\quad\quad\quad\quad\,\,\,\,\quad\quad+3\alpha_1/m_i^6+45/m_i^8\big\}{a^4\epsilon}/{\pi^4},\\
&\CT^{(-2)}_0(\vect n_i)\equiv-2au_0/\pi\epsilon^2n_i^2,\\
&\CT^{(-2)}_1(\vect n_i)\equiv-2\big[\widetilde{\rho}_{AA1}^{}(\vect n_i)+8\pi^3w\ln(\epsilon\lvert a\rvert)\big]{a^4}/{\pi^4\epsilon^2}\notag\\
&\quad\quad\quad\quad\quad\,\,-D/12\pi^2\epsilon^2,\\
&\CT^{(-2)}_2(\vect n_i)\equiv-6\big[\rho_{A1}^{}(\vect n_i)+2\alpha_1^{}/n_i^2-3/n_i^4\big]{a^4}/{\pi^4\epsilon^2},\\
&\CT^{(-2)}_3(\vect n_i)\equiv18a^4/\pi^4\epsilon^2n_i^2.
\end{align}
\end{subequations}
In preparation for the next step, we derive
\begin{align}
&t'^{(1,-3)}_{\vect q,\vect k}=X^{(-2)}_{\vect q,\vect k}J(\vect q)+\big\{96\pi w\alpha_1^{}a^4 Z_{1/\lvert a\rvert}(k)/k^3\notag\\
&+\big[-({96}w\alpha_1^{}/{\pi})a^4\ln(\epsilon\lvert a\rvert)+C_1'a^4-\alpha_1D/4\pi^2\notag\\
&\quad\,\,\,\,-3\alpha_1^{}a^3r_s^{}\big](2\pi)^3\delta(\vect k)\big\}\epsilon(2\pi\epsilon)^3\delta(\vect q),\\
&C_1'\equiv\pi^{-4}\big[-4\widetilde{\alpha}_{1AA1}^{}-48\pi^3w\alpha_1^{}\ln(2\pi)-\pi^3\alpha_1^{}C_0^{}\notag\\
&\quad\quad-12\alpha_{2A1}^{}-9\alpha_1^2\alpha_2^{}-6\alpha_2^2+3\alpha_1^{}\alpha_3^{}+45\alpha_4^{}\big],
\end{align}
where
\begin{align}
&\widetilde{\alpha}_{1AA1}^{}\equiv\lim_{N\rightarrow\infty}\big\{\big[\!\!\sum_{\vect m\ne0;m<N}m^{-2}\widetilde{\rho}_{AA1}^{}(\vect m)\big]\notag\\
&\quad\quad-32\pi^4w N\big[\ln(2\pi N)-1\big]+12\pi^3w\alpha_1^{}\ln N\big\}
\end{align}
is a 3-fold lattice sum (evaluated in Appendix~\ref{sec:constants}).

\textbf{\textit{Step 9.}}
From
\begin{align*}
&(H\CS^{(1)\vect q})_\vect k+\big[24\alpha_1a^2\epsilon^4\phi_\vect k-{3}\alpha_1a\epsilon^4W^{(0)}_\vect k/{\pi}\big](2\pi)^3\delta(\vect q)\\
&+\!\big\{6\big[n^2\rho_{A1}^{}(\vect n)\!+\!\alpha_1\!+\!n^{-2}\big]a^2\epsilon\phi_\vect k
\!-\!{3}\alpha_1a\epsilon W^{(0)}_\vect k\!/\!{\pi}\big\}J(\vect q)\!=\!0
\end{align*}
and [note that $t'^{(1,-7)}_{\vect q,\vect k}=t'^{(1,-6)}_{\vect q,\vect k}=0$]
\begin{equation*}
\CS^{(1)\vect q}_\vect k=\sum_{s=-5}^{-3}t'^{(1,s)}_{\vect q,\vect k}+O(\epsilon^1k^{-2})~~~(\text{for small }\vect k),
\end{equation*}
we get
\begin{align}
&\CS^{(1)\vect q}_\vect k=\big\{\big[-({96}w\alpha_1^{}/{\pi})a^4\ln(\epsilon\lvert a\rvert)+C_1a^4-{\alpha_1D}/{4\pi^2}\notag\\
&\quad-6\alpha_1a^3r_s+{6}\alpha_1au_0/{\pi}\big]\phi_\vect k-6\alpha_1a^2f_\vect k-{3}\alpha_1ad_\vect k/{\pi}\big\}\notag\\
&\quad\quad\quad\quad\quad\quad\quad\times\epsilon(2\pi\epsilon)^3\delta(\vect q)+Y(\vect q,\vect k)J(\vect q),
\end{align}
where
\begin{align}
C_1&\equiv C_1'+6w\alpha_1(7/\sqrt{3}-8/\pi)\notag\\
&=810.053~286~803~649~420,
\end{align}
and the actual formula for $Y(\vect q,\vect k)$ is not needed in this paper.

\textbf{\textit{Step 10.}}
Substituting the latest results for $\CS^{\vect q}_\vect k$ and $\CR_{\vect q_1\vect q_2\vect q_3}$ into Eq.~\eqref{eq:CT},
extracting all the terms that contain the factor $(2\pi)^6\delta(\vect q_1)\delta(\vect q_2)$ from this equation,
noting that $\CT_{\vect q_1\vect q_2\vect q_3}-(2\pi)^6\delta(\vect q_1)\delta(\vect q_2)$ vanishes at $\vect q_1=\vect q_2=0$, and
using Eqs.~\eqref{eq:intUphi} and \eqref{eq:D_complicated} to simplify the result, we get
\begin{align}
E&=12\pi a\epsilon^3\big\{1-\alpha_1a\epsilon/\pi+(\alpha_1^2+\alpha_2)a^2\epsilon^2/\pi^2\notag\\
&\quad+\big[16wa^3\ln(\epsilon\lvert a\rvert)+C_0a^3+3\pi a^2r_s\big]\epsilon^3\notag\\
&\quad+\big[-({96}w\alpha_1/{\pi})a^4\ln(\epsilon\lvert a\rvert)+C_1a^4-6\alpha_1a^3r_s\big]\epsilon^4\big\}\notag\\
&\quad\quad\quad\quad\quad\quad+\big(\epsilon^6-{6}\alpha_1^{}a\epsilon^7/{\pi}\big)D+O(\epsilon^8).
\end{align}
Substituting the numerical values of the lattice sums (see Appendix~\ref{sec:constants})
to the above equation, we get Eq.~\eqref{eq:Enumerical}. 

We will not proceed to determine $\CT^{(-1)}_{\vect q_1\vect q_2\vect q_3}$ in this paper.

\subsection{Results for Wu's parameter $\mathcal{E}_3$\label{subsec:E3}}
Wu computed the three-boson energy in the large periodic cubic volume to order $L^{-6}$, and he left an unknown parameter $\mathcal{E}_3$
for the three-boson interaction strength at low energy \cite{Wu}. $\mathcal{E}_3$ is needed to determine the full order-$\rho^2$ correction to
the many-body energy \cite{Wu}.

Comparing our Eq.~\eqref{eq:E_L-6} with Wu's result, Eq.~(5.29) of Ref.~\cite{Wu} (note that the unit of mass in \cite{Wu} is $2m_\text{boson}=1$),
we find that Huang and Yang's constant \cite{HuangYangExpansion} $C=-\alpha_1/\pi=2.837\cdots$, \cite{footnote:alpha_difference}
different from the number $2.37$ which was first provided in Ref.~\cite{HuangYangExpansion} and then adopted by \cite{Wu};
the symbols $\xi_s$ in Ref.~\cite{Wu} equal $\alpha_s/\pi^s$; the symbol $\mathcal{E}_3$ in Ref.~\cite{Wu} is
now expressed in terms of $D$ of the present paper:
\begin{align}
\mathcal{E}_3&=D/12\pi a^4+3\pi r_s/a\notag\\&\quad-4\alpha_{1A1}^{}/\pi^3+16w\ln(2\pi)-(14\pi/\sqrt{3}-16)w\notag\\
&=D/12\pi a^4+3\pi r_s/a+73.699~808~371~935~4035.\label{eq:E3}
\end{align}

For hard-sphere bosons, $D$ is given by Eq.~\eqref{eq:DHS} and $r_{s}=2a/3$, so
\begin{equation}\label{eq:E3HS}
\mathcal{E}_{3}=126.709~37\pm0.000~06~~~\text{(for hard spheres)}.
\end{equation}



\subsection{Results for the ground state wave function}
There are 6 expansion formulas for $A_{\vect n_1\vect n_2\vect n_3}$, each of which is valid in a sub-region of the
discrete momentum configuration space:
\begin{subequations}\label{eq:Aresult}
\begin{equation}
A_{000}\equiv L^6,
\end{equation}
\begin{widetext}
\begin{multline}
A_{0,\vect n,-\vect n}=-aL^5/\pi n^2+\big(\alpha_1/n^2+1/n^4\big)a^2L^4/\pi^2
+\big\{-\big[\alpha_1^2+\alpha_2+4\rho_{A1}^{}(\vect n)\big]/n^2+2\alpha_1/n^4+15/n^6\big\}a^3L^3/\pi^3+u_0L^3\\
+\big\{\big[-4\widetilde{\rho}_{AA1}^{}(\vect n)/n^2+48\pi^3wn^{-2}\ln({L}/{\lvert a\rvert})-\pi^3C_0/n^2
-(9\alpha_1^2+6\alpha_2)/n^4-12\rho_{A1}^{}(\vect n)/n^4+3\alpha_1/n^6+48/n^8\big]{a^4}/{\pi^4}\\
-\big(\alpha_1+n^{-2}\big)au_0/\pi-D/4\pi^2n^2-3a^3r_s/n^2\big\}L^2+O(L^1),
\end{multline}
\begin{multline}
A_{\vect n_1\vect n_2\vect n_3}=\!\sum_{i=1}^3\!\big\{2a^2L^4/\pi^2n_i^2
-2\big[\rho_{A1}^{}(\vect n_i)+\alpha_1/n_i^2-3/n_i^4\big]a^3L^3/\pi^3-DL^2/12\pi^2
+2\big[8\pi^3w\ln({L}/{\lvert a\rvert})-\widetilde{\rho}_{AA1}^{}(\vect n_i)\big]a^4L^2/\pi^4\big\}\\
\times G_{n_1n_2n_3}+\big\{6a^3L^3/\pi^3n_i^2-6\big[\rho_{A1}^{}(\vect n_i)
+2\alpha_1/n_i^2-3/n_i^4\big]a^4L^2/\pi^4\big\}G_{n_1n_2n_3}^2+18a^4L^2G_{n_1n_2n_3}^3/\pi^4n_i^2-2au_0L^2/\pi n_i^2\\+O(L^1),
\end{multline}
\begin{multline}
A_{0,\vect N,-\vect N}=\big\{L^3-\alpha_1aL^2/\pi+(\alpha_1^2+\alpha_2)a^2L/\pi^2-16wa^3\ln({L}/{\lvert a\rvert})+C_0a^3+3\pi a^2r_s-3u_0
+\big[({96}w\alpha_1/{\pi})a^4\ln({L}/{\lvert a\rvert})+C_1a^4\\
-\alpha_1D/4\pi^2-6\alpha_1a^3r_s+6\alpha_1au_0/\pi\big]L^{-1}\big\}\phi_{\vect k}+\big(1-3\alpha_1a/\pi L\big)d_\vect k
+\big(6\pi a-6\alpha_1a^2/L\big)f_\vect k+O(L^{-2}),
\end{multline}
\begin{multline}
A_{\vect n,-\vect n/2+\vect N,-\vect n/2-\vect N}=\big\{-2aL^2/\pi n^2
+\big[2\rho_{A1}^{}(\vect n)+2\alpha_1/n^2-6/n^4\big]a^2L/\pi^2-16wa^3\ln({L}/{\lvert a\rvert})
+2a^3\widetilde{\rho}_{AA1}^{}(\vect n)/\pi^3\big\}\phi_\vect k\\
+10\pi a\phi^{(d)}_{\hat{\vect n}\vect k}+d_\vect k+O(L^{-1}),
\end{multline}
\begin{equation}\label{eq:ANNNresult}
A_{\vect N_1\vect N_2\vect N_3}=\big[1-3\alpha_1a/\pi L+3(2\beta_{1A}^{}+\alpha_1^2-3\alpha_2)a^2/\pi^2L^2\big]
\phi^{(3)}_{\vect k_1\vect k_2\vect k_3}+O(L^{-3}),
\end{equation}
\end{widetext}
\end{subequations}
where $\vect n$'s are \emph{nonzero} vectors of order unity, $\vect N$'s are large vectors of order $L/\max(r_e,\lvert a\rvert)$
($r_e$ is the range of the interaction),
$\vect k\equiv2\pi\vect N/L$, $\vect k_i\equiv2\pi\vect N_i/L$, and $G_{n_1n_2n_3}\equiv2/(n_1^2+n_2^2+n_3^2)$.
The formulas for $A_{0,0,0}$, $A_{0,\vect n,-\vect n}$, and $A_{\vect n_1\vect n_2\vect n_3}$ are extracted from the above results for
$\CT_{\vect q_1\vect q_2\vect q_3}$; those for $A_{0,\vect N,-\vect N}$ and $A_{\vect n,-\vect n/2+\vect N,-\vect n/2-\vect N}$
are extracted from $\CS^{\vect q}_\vect k$; the formula for $A_{\vect N_1\vect N_2\vect N_3}$ results from the expansion
of $\CR_{\vect k_1\vect k_2\vect k_3}$. The numerical constants in the above formulas are computed in Appendix~\ref{sec:constants}.

\subsection{Momentum distribution\label{subsec:momentum}}
The expectation value of the number of bosons with momentum $2\pi\vect n/L$ is
$
\mathcal{N}_\vect n=c\sum_{\text{all }\vect n'}\lvert A_{\vect n,\vect n',-\vect n-\vect n'}\rvert^2
$
for some constant $c$ such that $\sum_{\text{all }\vect n}\mathcal{N}_\vect n=3$.

For any nonzero integral vector $\vect n$ of order unity and any large integral vector $\vect N$ of order $L/\max(r_e,\lvert a\rvert)$, we have
\begin{subequations}
\begin{equation}
\mathcal{N}_0=c\lvert A_{000}\rvert^2+c\!\!\!\!\sum_{0<n'\le N_c}\!\!\!\!\lvert A_{0,\vect n',-\vect n'}\rvert^2
+c\!\!\sum_{N'>N_c}\!\!\lvert A_{0,\vect N',-\vect N'}\rvert^2,
\end{equation}
\begin{align}
\mathcal{N}_\vect n&=2c\lvert A_{0,\vect n,-\vect n}\rvert^2+c\!\!\!\!\sum_{\vect n'\ne\pm\vect n/2,~n'\le N_c}\!\!\!\!
\lvert A_{\vect n,-\vect n/2+\vect n',-\vect n/2-\vect n'}\rvert^2\notag\\
&\quad+c\sum_{N'>N_c}\lvert A_{\vect n,-\vect n/2+\vect N',-\vect n/2-\vect N'}\rvert^2,\label{eq:Nn}
\end{align}
\begin{align}
\mathcal{N}_\vect N&=2c\lvert A_{0,\vect N,-\vect N}\rvert^2+2c\sum_{\vect n'\ne0,~n'\le N_c}\lvert A_{\vect n',\vect N,-\vect n'-\vect N}\rvert^2
\notag\\
&\quad+c\sum_{N'>N_c,~\lvert \vect N'+\vect N\rvert>N_c}\lvert A_{\vect N',\vect N,-\vect N'-\vect N}\rvert^2,
\end{align}
\end{subequations}
where $N_c$ satisfies $1\ll N_c\ll L/\max(r_e,\lvert a\rvert)$; in Eq.~\eqref{eq:Nn}, $\vect n'\pm\vect n/2$ and $\vect N'\pm\vect n/2$ have
integral Cartesian components. Substituting Eqs.~\eqref{eq:Aresult} and using Eq.~\eqref{eq:IntPhikSquare}, we get
\begin{align*}
\mathcal{N}_0/c&=L^{12}+\alpha_2a^2L^{10}/\pi^2-2(\alpha_1\alpha_2+\alpha_3)a^3L^9/\pi^3\notag\\&\quad-2\Re u_0L^9-2\pi a^2r_sL^9+O(L^8),\\
\mathcal{N}_\vect n/c&=2a^2L^{10}/\pi^2n^4-4\big(\alpha_1/n^4+1/n^6\big)a^3L^9/\pi^3\notag\\
&\quad+O(L^8),\\
\mathcal{N}_\vect N/c&=2L^6\big(1-2\alpha_1a/\pi L\big)\lvert\phi_{2\pi\vect N/L}^{}\rvert^2+O(L^4),
\end{align*}
Solving $\sum_{\text{all }\vect n}\mathcal{N}_\vect n=3$ [using Eq.~\eqref{eq:IntPhikSquare} again], we get
\begin{align}
c&=3L^{-12}\big[1-3\alpha_2a^2/\pi^2L^2+6(\alpha_1\alpha_2+\alpha_3)a^3/\pi^3L^3\notag\\
&\quad\quad\quad\quad+6\Re u_0/L^3+6\pi a^2r_s/L^3+O(L^{-4})\big],
\end{align}
and
\begin{widetext}
\begin{subequations}
\begin{align}
\mathcal{N}_0/3-1&=-2\alpha_2a^2/\pi^2L^2+4(\alpha_1\alpha_2+\alpha_3)a^3/\pi^3L^3+(4\Re u_0+4\pi a^2r_s)/L^3+O(L^{-4})\notag\\
&=-3.350147643\,a^2/L^2+(4\Re u_0+4\pi a^2r_s-17.926831164\,a^3)/L^3+O(L^{-4}),\label{eq:depletion3}
\end{align}
\begin{align}
\mathcal{N}_\vect n&=6a^2/\pi^2L^2n^4-(12a^3/\pi^3L^3)\big(\alpha_1/n^4+1/n^6\big)+O(L^{-4})\notag\\
&=(0.607927102/n^4)({a}/{L})^2+\big(3.449740068/n^4-0.387018413/n^6\big)({a}/{L})^3+O(L^{-4}),
\end{align}
\begin{align}
\mathcal{N}_\vect N&=6L^{-6}\big(1-2\alpha_1a/\pi L\big)\lvert\phi_{2\pi\vect N/L}^{}\rvert^2+O(L^{-8})
=6L^{-6}\big(1+5.674594959\,a/L\big)\lvert\phi_{2\pi\vect N/L}^{}\rvert^2+O(L^{-8}).
\end{align}
\end{subequations}
\end{widetext}
One can derive higher order results for $\mathcal{N}$'s from Eqs.~\eqref{eq:Aresult}.

Equation~\eqref{eq:depletion3} shows that the depletion of the population of the zero momentum state
depends on the parameter $u_0$ at the next-to-leading order in the volume expansion.

Given $\phi(\vect r)$, one can compute $u_0$ and $r_s$ from the formulas
\begin{subequations}\label{eq:u0rs}
\begin{align}
u_0&=\int\big[\phi(\vect r)-(1-a/r)\big]\dif^3r,\label{eq:u0}\\
-2\pi a^2r_s&=\int\big[\lvert\phi(\vect r)\rvert^2-(1-a/r)^2\big]\dif^3r.
\end{align}
\end{subequations}
Because $\phi(\vect r)=1-a/r$ outside of the range of the interaction, the integrands are nonzero within the range only.
Because there exists a two-body potential $V(\vect r)=\phi^{-1}(\vect r)\nabla^2\phi(\vect r)$ for any function $\phi(\vect r)$,
Eqs.~\eqref{eq:u0rs} indicate that $u_0$ is in general \emph{independent} from $a$ and $r_s$.

After repeated failure to find a general relation between $\Re u_0$ and any finite number of parameters in Eq.~\eqref{eq:phaseshift},
the author conjectures that such a relation does not exist at all \cite{footnote:u0independence}.
Because the EFT \cite{Braaten2001} is formulated in terms of parameters in Eq.~\eqref{eq:phaseshift} along with
3-body, 4-body, \dots parameters, the \emph{two-body} parameter $u_0$ is absent in the EFT.

\subsection{Generalization to $\CN$ bosons}
\subsubsection{Results}
To understand the ground state energy and momentum distribution of dilute Bose-Einstein condensates (BECs),
the author generalized the above calculations to $\CN$ bosons ($\CN=1,2,3,4\dots$).

Solving the $\CN$-boson Schr\"{o}dinger equation perturbatively in powers of $1/L$, using the same \emph{finite-range} interactions
as above, the author obtained the volume expansion for $E$ up to the order $L^{-6}$, and found that it exactly agrees with
Beane, Detmold, and Savage's result for $E$ \cite{Savage} which was derived with zero-range pseudopotentials \cite{Savage}, if
the three-boson contact interaction parameter $\eta_3(\mu)$ in Ref.~\cite{Savage} satisfies
\begin{align}
&\eta_3(\lvert a\rvert^{-1})-48a^4(4\mathcal{Q}+2\mathcal{R})/\pi^2\notag\\
&=D+12\pi^2a^3r_s-48a^4\alpha_{1A1}/\pi^2\notag\\&\quad+24\pi w a^4\big[8\ln(2\pi)-(7\pi/\sqrt{3}-8)\big]\notag\\
&=D+12\pi^2a^3r_s+2778.417~318~626~973~645a^4,
\end{align}
where $4\mathcal{Q}+2\mathcal{R}$ is a number defined in Ref.~\cite{Savage}.
It is the combination $\eta_3(\mu)-48a^4(4\mathcal{Q}+2\mathcal{R})/\pi^2$
that appears in their formula \cite{Savage} for $E$. In summary,
\begin{align}
&E=P(1/L;a,\CN)-\frac{192\pi w a^4}{L^6}\binom{\CN}{3}\ln\frac{L}{\lvert a\rvert}\notag\\
&+\binom{\CN}{3}\frac{D+12\pi^2a^3r_s}{L^6}+\binom{\CN}{2}\frac{8\pi^2a^3r_s}{L^6}+O(L^{-7}),
\end{align}
where $P$ is a well-determined \cite{Savage} power series in $1/L$ with $a$ and $\CN$ as the only parameters, and
$\binom{n}{i}=\frac{n!}{i!(n-i)!}$.

In addition to the energy, the present author obtained the following results for the momentum distribution:
\begin{subequations}\label{eq:CN}
\begin{align}
&x\equiv\CN_0/\CN-1=-(\CN-1)\alpha_2(a/\pi L)^2\notag\\
&+2(\CN-1)(\Re u_0+\pi a^2r_s)/L^3\notag\\
&+2(\CN-1)\big[\alpha_1\alpha_2+(2\CN-5)\alpha_3\big](a/\pi L)^3+O(L^{-4}),\label{eq:x}
\end{align}
\begin{align}
\CN_{\vect n}&=\CN(\CN-1)n^{-4}(a/\pi L)^2\notag\\
&\quad-2\CN(\CN-1)\big[\alpha_1/n^4+(2\CN-5)/n^6\big](a/\pi L)^3\notag\\
&\quad+O(L^{-4}),\label{eq:CNn}
\end{align}
\begin{equation}
\CN_{\vect N}=\CN(\CN-1)L^{-6}(1-2\alpha_1a/\pi L)\lvert\phi_{2\pi\vect N/L}\rvert^2+O(L^{-8}).\label{eq:CNN}
\end{equation}
\end{subequations}
Equations~\eqref{eq:CN}, unlike $E$, can \emph{not} be derived from the zero-range pseudopotentials or the effective field theory.
[Even in Eq.~\eqref{eq:CNn}, the $L^{-4}$ correction will contain the short-range parameter $u_0$.]

\subsubsection{Implications for dilute Bose-Einstein condensates
in the thermodynamic limit\label{subsubsec:EAndx}}

Let $a>0$. Let $\rho=\CN/L^3$. At large $\CN$ there are two different low-density regimes (and an intermediate regime between the two),
depending on the box size $L$:
\begin{itemize}
\item $L\gg\CN a$, so that $L$ is \emph{small} compared to the BEC healing length $\xi\sim(\rho a)^{-1/2}$;
\item $\CN^{1/3}a\ll L\ll\CN a$, and the system is a dilute BEC near the thermodynamic limit
($L\gg\xi$).
\end{itemize}

The $1/L$ expansions for the energy and the momentum distribution
are valid in the first regime only. In the second regime, they diverge like
$\sum_{i=0}^{\infty}c_i(a\CN/L)^i$. Nevertheless, one can infer many properties of
the BEC in the second regime.

Now we tentatively take the thermodynamic limit ($\rho$ fixed, $\CN$ and $L$ large)
of the $1/L$ expansions of $E_0=E/\CN$ and $x$ [see Eq.~\eqref{eq:x}].
Each term that remains finite is retained;
each term that diverges must be rendered finite by a resummation \cite{LeeHuangYang1957}
that includes all similar, but higher order (and increasingly more divergent) contributions \cite{LeeHuangYang1957}.

Thus the mean-field energy term for $E_0$ is reproduced. The logarithmic term
$\approx-32\pi w a^4\rho^2\ln(L/a)$ is rendered finite by a resummation that
changes $L$ to $O(\xi)$, yielding precisely the same logarithmic term as in Eq.~\eqref{eq:E_BEC}.
The leading nonuniversal terms (in the sense that parameters other than $a$ contribute)
become $\rho^2(D/6+2\pi^2a^3r_s)$. Comparing these findings with Eq.~\eqref{eq:E_BEC}, we infer that
\begin{align}
E_0&=2\pi\rho a\big[1+(128/15\sqrt{\pi})(\rho a^3)^{1/2}+8w\rho a^3\ln(\rho a^3)\notag\\
&\quad\quad\quad\quad\mspace{7mu}+(D/12\pi a^4+\pi r_s/a+C^E)\rho a^3\big]
\label{eq:E0_thermodynamic}
\end{align}
plus higher order terms in density $\rho$, where $C^E$ is a universal constant that
remains the same for all Bose gases. Because $E_0$ was computed by Braaten and Nieto \cite{Braaten1999},
a comparison will be made between Eq.~\eqref{eq:E0_thermodynamic} and Ref.~\cite{Braaten1999}
(in Sec.~\ref{subsec:BECenergy}).

The contributions to $x$ that depend on $a$ only are divergent in the thermodynamic limit. After a resummation that includes
higher order terms in $\CN a/L$ (analogous to Eq.~(55) of Ref.~\cite{LeeHuangYang1957}),
they must reproduce Bogoliubov's well-known formula $x=-\frac{8}{3\sqrt{\pi}}\sqrt{\rho a^3}+\dots.$
The leading nonuniversal term becomes $\rho(2\Re u_0+2\pi a^2r_s)$; because this term
comes from the short-range behavior in two-body collisions, it must extend into the thermodynamic limit.
Combining these observations with an EFT prediction for the condensate depletion \cite{Braaten2001}
(which should be valid through order $\rho^1$ at $u_0=a^2r_s=0$, at the very least), we infer that
\begin{equation}\label{eq:x_thermodynamic}
x=-\frac{8}{3\sqrt{\pi}}\sqrt{\rho a^3}+\rho(2\Re u_0+2\pi a^2r_s)+C^x\rho a^3
\end{equation}
plus higher order terms in $\rho$, where $C^x$ is another universal numerical constant.
Thus the nonuniversal effect in the quantum depletion of the condensate is larger than the EFT prediction \cite{Braaten2001}
by a factor of order $(\rho a^3)^{-1/2}$ at low density, if $u_0^{1/3}\sim r_s\sim a$.
This disagreement is entirely caused by the fact that the momentum distribution at $k\sim1/r_e\gg(\rho a)^{1/2}$
is approximately $\rho^2\lvert\phi_\vect k\rvert^2$,
rather than a structureless function $\approx16\pi^2a^2\rho^2/k^4$ as implied by the EFT.

The condensate fraction (namely $\CN_0/\CN$, or $1+x$) of a dilute Bose gas of hard spheres
(for which $2u_0+2\pi a^2r_s=\frac{8\pi}{3}a^3$)
is thus slightly \emph{greater} than that of a Bose gas with $r_e\ll a$ [for which $\lvert 2u_0+2\pi a^2r_s\rvert\ll a^3$]
by about $\frac{8\pi}{3}\rho a^3$ at zero temperature,
if the two gases have the same number density and scattering length.

\section{Low energy scattering amplitudes of three identical bosons\label{sec:Tmat}}

In this section we compute the T-matrix elements of three identical bosons
at low energy. We will compare our results with a similar calculation in Ref.~\cite{Braaten1999},
to establish the relation between the parameter $D$ in the present paper and the three-body contact
interaction parameter $g_3(\kappa)$ in \cite{Braaten1999}.

The interactions are the same as in Sec.~\ref{sec:asymp}.

The T-matrix is denoted with roman type $\Tmat$ below, since the italic $T$ is already used for the wave function components.

\subsection{2-boson scattering amplitude}

The two-boson T-matrix can be expressed in terms of the scattering phase shifts \cite{footnote:CollisionTheory} ($\hbar=m_\text{boson}=1$):
\begin{equation}\label{eq:Tmat2exact}
\Tmat(b,\hat{\vect b}\cdot\hat{\vect q})
=(4\pi/\I q)\!\!\sum_{l=0,2,4,\dots}\!\!\big[\e^{2\I\delta_l(q)}-1\big](2l+1)P_l(\hat{\vect b}\cdot\hat{\vect q}),
\end{equation}
where $\pm\vect b$ and $\pm\vect q$ are the momenta of the two bosons before and after the scattering, respectively,
and $q=b$.

At small $q$ we apply Eq.~\eqref{eq:phaseshift} to the above formula and get
\begin{align}
\Tmat(b,\hat{\vect b}\cdot\hat{\vect q})&=-8\pi a+\I\,8\pi a^2q+8\pi a^2(a-r_s/2)q^2\notag\\&\quad-\I\,8\pi a^3(a-r_s)q^3
+O(q^4).\label{eq:Tmat2approx}
\end{align}
The first two terms in this expansion agree with Ref.~\cite{Braaten1999}, and all higher order corrections disagree,
since the effective range $r_s$ is not included in \cite{Braaten1999}.
Equation~\eqref{eq:Tmat2approx} agrees with Ref.~\cite{Braaten2001} where $r_s$ is taken into account.

\subsection{3-boson scattering amplitude\label{sec:T3}}
Equation~\eqref{eq:Tmat2approx} can be alternatively derived from a systematic perturbative solution to the two-body Schr\"{o}dinger equation
at incoming momenta $\pm\vect b$ and energy $E=b^2$:
\begin{align*}\Psi_\vect q&=\frac{1}{2}(2\pi)^3\big[\delta(\vect q+\vect b)+\delta(\vect q-\vect b)\big]
+\frac{\Tmat(b,\hat{\vect b}\cdot\hat{\vect q})}{2(q^2-E-\I\eta)}
\\&\quad+\text{(terms that are regular at $q^2=E$)},
\end{align*}
where $-\I\eta$ specifies an outgoing wave ($\eta\rightarrow0^+$).

Similarly, from the stationary wave function $\Psi_{\vect q_1\vect q_2\vect q_3}$
describing the scattering of three bosons with incoming momenta $\vect b_1$, $\vect b_2$, and $\vect b_3$,
and energy
\begin{equation}E=(b_1^2+b_2^2+b_3^2)/2,
\end{equation}
one can extract the three-boson T-matrix elements:
\begin{align}
&\Psi_{\vect q_1\vect q_2\vect q_3}=\frac{(2\pi)^6}{6}\sum_{P}\delta(\vect q_1-\vect b_{P1})\delta(\vect q_2-\vect b_{P2})\notag\\&
+\frac{1}{6}G^E_{q_1q_2q_3}\Big[\Tmat(\vect b_1\vect b_2\vect b_3;\vect q_1\vect q_2\vect q_3)\notag\\&\mspace{93mu}+\sum_{i,j=1}^3
\Tmat(h_i,\hat{\vect h_i}\cdot\hat{\vect p_j})(2\pi)^3\delta(\vect q_j-\vect b_i)\Big]
\notag\\&+\text{[terms that are regular at $(q_1^2+q_2^2+q_3^2)/2=E$]}.\label{eq:PsiForm}
\end{align}
Here $P$ refers to all 6 permutations of ``123", $\sum_{i=1}^3\vect b_i\equiv\sum_{i=1}^3\vect q_i\equiv0$,
$\vect p_i$ is defined in Eq.~\eqref{eq:p}, and
\begin{equation}
G^E_{q_1q_2q_3}=\big[(q_1^2+q_2^2+q_3^2)/2-E-\I\eta\big]^{-1},
\end{equation}
\begin{equation}
\vect h_1=(\vect b_2-\vect b_3)/2, \text{~and similarly for $\vect h_2$, $\vect h_3$}.
\end{equation}

Let $b_i\sim q_i\sim q$ be small. We determine $\Psi$ perturbatively next.

The equations for $\Psi$ here are formally identical with Eqs.~\eqref{eq:CRCSCT} in Sec.~\ref{sec:cubic}, except for three differences:
1) the box size $L=\infty$ here, so $I(\vect q)\equiv1$, 2) $E\sim q^2$ here, instead of $1/L^3$ in Sec.~\ref{sec:cubic}, and
3) the leading contribution to $\Psi_{\vect q_1\vect q_2\vect q_3}$ is
\begin{equation}
T^{(-6)}_{\vect q_1\vect q_2\vect q_3}=\frac{(2\pi)^6}{6}\sum_{P}\delta(\vect q_1-\vect b_{P1})\delta(\vect q_2-\vect b_{P2}).
\end{equation}

Naturally, $\Psi_{\vect q_1\vect q_2\vect q_3}\rightarrow\phi^{(3)}_{\vect q_1\vect q_2\vect q_3}$ in the limit $E, b_i\rightarrow0$,
and the calculation here will be a generalization of Sec.~\ref{sec:asymp} to nonzero incoming momenta. So we use
the same symbols $T$ ($\ne\Tmat$)
and $S$ as in Sec.~\ref{sec:asymp}, in the asymptotic expansions
\begin{gather}
\Psi_{\vect q_1\vect q_2\vect q_3}=\sum_{s=-6}^\infty T^{(s)}_{\vect q_1\vect q_2\vect q_3},\\
\Psi^\vect q_\vect k\equiv\Psi_{\vect q,-\vect q/2+\vect k,-\vect q/2-\vect k}=\sum_{s=-3}^\infty S^{(s)\vect q}_\vect k,
\end{gather}
where $T^{(s)}_{\vect q_1\vect q_2\vect q_3}$ and $S^{(s)\vect q}_\vect k$ scale like $q^s$ (not excluding $q^s\ln^mq$).
When $E,b_i\rightarrow0$ but $\vect q$'s and $\vect k$ are fixed, the (complicated) results for $T^{(s)}$ and $S^{(s)}$
in this section will reduce to the much simpler ones in Sec.~\ref{sec:asymp}.

When the three bosons all come to a region of size $\sim r_e$ (radius of interaction), effects due to a nonzero $E$ are small.
So at momenta $k_i\sim 1/r_e\gg\sqrt{E}$ \cite{footnote:Psik1k2k3}
\begin{equation}\label{eq:Psik1k2k3}
\Psi_{\vect k_1\vect k_2\vect k_3}=\phi^{(3)}_{\vect k_1\vect k_2\vect k_3}+O(\sqrt{E}).
\end{equation}

Employing the same systematic expansion method as in Secs.~\ref{sec:asymp} and \ref{sec:cubic}, we obtain
the following results (listed in the same order as they were obtained):
\begin{subequations}
\begin{equation}
S^{(-3)\vect q}_\vect k=\frac{1}{3}\sum_{i=1}^3(2\pi)^3\delta(\vect q-\vect b_i)\phi_\vect k,
\end{equation}
\begin{equation}
T^{(-5)}_{\vect q_1\vect q_2\vect q_3}=-\frac{4\pi a}{3}G^E_{q_1q_2q_3}\sum_{i,j=1}^3(2\pi)^3\delta(\vect q_j-\vect b_i),
\end{equation}
\begin{equation}
S^{(-2)\vect q}_\vect k=-\frac{1}{3}\sum_{i=1}^3\big[8\pi aG^E_{\vect q\vect b_i}+\I a h_i(2\pi)^3\delta(\vect q-\vect b_i)\big]\phi_\vect k,
\end{equation}
\begin{equation}
T^{(-4)}_{\vect q_1\vect q_2\vect q_3}\!\!=\!\!\frac{16\pi^2a^2}{3}G^E_{q_1q_2q_3}\!\!\!
\sum_{i,j=1}^3\!\!\Big[2G^E_{\vect q_j\vect b_i}+\frac{\I h_i}{4\pi}(2\pi)^3\delta(\vect q_j-\vect b_i)\Big],
\end{equation}
\begin{widetext}
\begin{equation}
S^{(-1)\vect q}_\vect k\!\!=\!\frac{1}{3}\sum_{i=1}^3\!\Big\{\Big[8\pi a^2(\I h_i-\sqrt{3q^2/4-E-\I\eta})G^E_{\vect q\vect b_i}
+64\pi^2 a^2c^E_1(\vect q,\vect b_i)
\Big]\phi_\vect k+h_i^2(2\pi)^3\delta(\vect q-\vect b_i)\Big[f_\vect k-a(a-r_s/2)\phi_\vect k-5\phi^{(d)}_{\hat{\vect h}_i\vect k}\Big]\!\Big\},
\end{equation}
\begin{align}
T^{(-3)}_{\vect q_1\vect q_2\vect q_3}&\!=\!\frac{1}{3}G^E_{q_1q_2q_3}\!\!
\sum_{i,j=1}^3\!\big[32\pi^2a^3(\sqrt{3q_j^2/4-E-\I\eta}-\!\I h_i)G^E_{\vect q_j\vect b_i}
\!-\!256\pi^3a^3c^E_1(\vect q_j,\vect b_i)+4\pi a^2(a-r_s/2)h_i^2(2\pi)^3\delta(\vect q_j-\vect b_i)\big]\notag\\
&\quad+\frac{u_0}{3}\sum_{i,j=1}^3(2\pi)^3\delta(\vect q_j-\vect b_i),
\end{align}
\begin{align}
&S^{(0)\vect q}_\vect k=d_\vect k
-\frac{1}{3}\sum_{i=1}^3\Big[6\pi a+2\pi a(4E-3q^2)G^E_{\vect q\vect b_i}+\I ah_i^3(2\pi)^3\delta(\vect q-\vect b_i)\Big]f_\vect k
+\frac{10\pi a}{3}\sum_{i=1}^3{q'_i}^2G^E_{\vect q\vect b_i}\phi^{(d)}_{\hat{\vect q}'_i\vect k}\notag\\&+\frac{1}{3}\sum_{i=1}^3\Big[
64\pi^2a^3(\sqrt{3q^2/4-E-\I\eta}-\I h_i)c^E_1(\vect q,\vect b_i)-256\pi^3a^3c^E_2(\vect q,\vect b_i,\lvert a\rvert^{-1})
+8\pi a^2(a-r_s/2)(E-3q^2/4+h_i^2)G^E_{\vect q\vect b_i}\notag\\
&\mspace{70mu}+\I\,8\pi a^3h_i\sqrt{3q^2/4-E-\I\eta}G^E_{\vect q\vect b_i}-3\pi a^2r_s-(14\pi/\sqrt{3}-16)wa^3
+\I a^2(a-r_s)h_i^3(2\pi)^3\delta(\vect q-\vect b_i)\Big]\phi_\vect k,
\end{align}
\begin{align}
&T^{(-2)}_{\vect q_1\vect q_2\vect q_3}=G^E_{q_1q_2q_3}\Big\{-D-\frac{4\pi a}{3}\sum_{i,j=1}^3\Big[
64\pi^2a^3(\sqrt{3q_j^2/4-E-\I\eta}-\I h_i)c^E_1(\vect q_j,\vect b_i)-256\pi^3a^3c^E_2(\vect q_j,\vect b_i,\lvert a\rvert^{-1})\notag\\
&+8\pi a^2(a-r_s/2)(E-3q_j^2/4+h_i^2)G^E_{\vect q_j\vect b_i}+\I\,8\pi a^3h_i\sqrt{3q_j^2/4-E-\I\eta}G^E_{\vect q_j\vect b_i}
-3\pi a^2r_s-(14\pi/\sqrt{3}-16)wa^3\notag\\&+\I\,a^2(a-r_s)h_i^3(2\pi)^3\delta(\vect q_j-\vect b_i)\Big]\Big\}
+\frac{u_0}{3}\sum_{i,j=1}^3\big[-8\pi aG^E_{\vect q_j\vect b_i}-\I ah_i(2\pi)^3\delta(\vect q_j-\vect b_i)\big],
\end{align}
\end{widetext}
\end{subequations}
where $\I\ne i$, $\vect q'_i\equiv\vect q+2\vect b_i$ (\emph{not} the outgoing momenta), $\sqrt{z}$
is defined with a branch cut along the negative $z$-axis, and
\begin{equation}
G^E_{\vect q\vect b}\equiv(q^2+\vect q\cdot\vect b+b^2-E-\I\eta)^{-1},
\end{equation}
\begin{equation}
c^E_1(\vect q,\vect b)\equiv\int\frac{\dif^3k}{(2\pi)^3}G^E_{\vect q\vect k}G^E_{\vect k\vect b},
\end{equation}
\begin{align}
&c^E_2(\vect q,\vect b,\kappa)\equiv\lim_{K\rightarrow\infty}-\frac{w}{16\pi^3}\ln\frac{K}{\kappa}+
\int_{k<K}\frac{\dif^3k}{(2\pi)^3}G^E_{\vect q\vect k}\notag\\
&\times\Big[2c^E_1(\vect k,\vect b)-\frac{\sqrt{3k^2/4-E-\I\eta}}{4\pi}G^E_{\vect k\vect b}\Big].
\end{align}
The loop integrals $c^E_1$ and $c^E_2$ emerge from the $Z$-$\delta$ expansions (Appendix~\ref{sec:Z-delta})
of $T^{(-4)}_{\vect q,-\vect q/2+\vect k,-\vect q/2-\vect k}$
and $T^{(-3)}_{\vect q,-\vect q/2+\vect k,-\vect q/2-\vect k}$, respectively.

Remarkably, $c^E_2$ (or any loop integral or lattice sum encountered in the present paper) is free from uncontrolled ultraviolet divergence;
this finiteness follows naturally from the rules of the $Z$-$\delta$ expansion
(as can be easily understood from a similar but simpler problem, Example 2 in Appendix~\ref{sec:Z-delta}).

Comparing the above results for $\Psi_{\vect q_1\vect q_2\vect q_3}$ [$=\sum_sT^{(s)}_{\vect q_1\vect q_2\vect q_3}$] with
Eq.~\eqref{eq:PsiForm}, we find the same result for the two-boson T-matrix as Eq.~\eqref{eq:Tmat2approx}, and
the first 3 terms in the low-energy expansion of the three-boson T-matrix:
\begin{equation}
\Tmat(\vect b_1\vect b_2\vect b_3;\vect q_1\vect q_2\vect q_3)=\sum_{s=-2}^\infty\Tmat^{(s)}(\vect b_1\vect b_2\vect b_3;\vect q_1\vect q_2\vect q_3),
\end{equation}
Where $\Tmat^{(s)}(\vect b_1\vect b_2\vect b_3;\vect q_1\vect q_2\vect q_3)\sim E^{s/2}$ (\emph{not} excluding $E^{s/2}\ln^mE$).
\begin{subequations}\label{eq:T3}
\begin{equation}
\Tmat^{(-2)}(\vect b_1\vect b_2\vect b_3;\vect q_1\vect q_2\vect q_3)=64\pi^2a^2\sum_{i,j=1}^3G^E_{\vect q_j\vect b_i},
\end{equation}
\begin{align}
&\Tmat^{(-1)}(\vect b_1\vect b_2\vect b_3;\vect q_1\vect q_2\vect q_3)=\sum_{i,j=1}^3\big[-512\pi^3a^3c^E_1(\vect q_j,\vect b_i)\notag\\
&\mspace{150mu}-\I\mspace{2mu}64\pi^2a^3(p_j+h_i)G^E_{\vect q_j\vect b_i}\big],\label{eq:T3m1}
\end{align}
\begin{align}
&\Tmat^{(0)}(\vect b_1\vect b_2\vect b_3;\vect q_1\vect q_2\vect q_3)=-6\big[D-24\pi w(7\pi/\sqrt{3}-8)a^4\notag\\
&\mspace{10mu}-36\pi^2a^3r_s\big]+64\pi^2a^3\sum_{i,j=1}^3\big[32\pi^2ac^E_2(\vect q_j,\vect b_i,\lvert a\rvert^{-1})\notag\\
&\mspace{50mu}+(r_s/2-a)(p_j^2+h_i^2)G^E_{\vect q_j\vect b_i}-ap_jh_iG^E_{\vect q_j\vect b_i}\notag\\
&\mspace{50mu}+\I\mspace{3mu}8\pi a(p_j+h_i)c^E_1(\vect q_j,\vect b_i)\big].\label{eq:T3m0}
\end{align}
\end{subequations}
The momenta
$\vect q_i$'s satisfy the constraint $\sum_{i=1}^3q_i^2/2=E$, in addition to $\sum_{i=1}^3\vect q_i=\sum_{i=1}^3\vect b_i=0$,
in Eqs.~\eqref{eq:T3}.

The two lowest order contributions to the 3-boson T-matrix, as well as the logarithmic dependence on energy at the third order
[in $a^4c^E_2(\vect q_j,\vect b_i,\lvert a\rvert^{-1})$],
are universal in the sense that they depend on the scattering
length $a$ only, in agreement with Ref.~\cite{Amado}. The leading nonuniversal contributions are
\begin{equation*}
-6(D-36\pi^2a^3r_s)+32\pi^2a^3r_s\sum_{i,j=1}^3(p_j^2+h_i^2)G^E_{\vect q_j\vect b_i}.
\end{equation*}
The momentum-independent term in this expression is proportional to $D-36\pi^2a^3r_s$,
while the leading nonuniversal contribution to the BEC energy
[in Eq.~\eqref{eq:E0_thermodynamic}] is proportional to $D+12\pi^2a^3r_s$.
We conclude that they can \emph{not} be absorbed into a single 3-body contact interaction parameter $g_3$,
\emph{unless} the two-body effective range $r_s=0$. This disagrees with Ref.~\cite{Braaten1999}.

\subsection{Comparison with Ref.~\cite{Braaten1999}}
If $r_s=0$, Eqs.~\eqref{eq:T3} then agree with Ref.~\cite{Braaten1999}. To see this, we use the power series for the 2-boson
T-matrix to expand the last term in Eq.~(77) of \cite{Braaten1999} (note that $q_{12}$ in \cite{Braaten1999} is twice $h_{3}$
here), and get all the 4 terms in Eqs.~\eqref{eq:T3} of the present paper which contain $G^E_{\vect q_j\vect b_i}$. The quantity
$\mathcal{T}_1^\text{1PI}$ of \cite{Braaten1999} corresponds to the first term on the right hand side of Eq.~\eqref{eq:T3m1}
above; the finite contribution to $\mathcal{T}_2^\text{1PI}$ in Eq.~(80) of \cite{Braaten1999} corresponds to
the last term of Eq.~\eqref{eq:T3m0} above. The sum of the divergent \cite{Braaten1999}, but dimensionally regularized \cite{Braaten1999},
terms in $\mathcal{T}_2^\text{1PI}$ and the 3-body term $-\big[g_3(\kappa)+\delta g_3(\kappa)\big]$ \cite{Braaten1999}
can be expressed in terms of $c^E_2$ defined above:
$$\frac{(8\pi a)^4}{2}\Big[c_\text{MS}+\sum_{i,j=1}^3c^E_2(\vect k_i,\vect k_j',\kappa)\Big]-g_3(\kappa),$$
where
\begin{align*}
c_\text{MS}&=\frac{18w\big[\ln(2\pi)+1-\gamma\big]+\sqrt{3}\big[2\delta'+9\ln 3-18\big]}{64\pi^3}\\
&=0.055~571~793~27.
\end{align*}
Here $\delta'=\sum_{n=0}^\infty\big[(1/3+n)^{-2}-(2/3+n)^{-2}\big].$
The momentum-dependent terms in Eqs.~\eqref{eq:T3} thus completely agree with Ref.~\cite{Braaten1999} at $r_s=0$.
Further matching the constant terms, we find the relation between the 3-body parameter $g_3(\kappa)$ of Ref.~\cite{Braaten1999}
and the scattering hypervolume $D$ defined in the present paper at $r_s=0$:
\begin{align*}
g_3(\lvert a\rvert^{-1})&=6\big[D-24\pi w(7\pi/\sqrt{3}-8)a^4\big]+\frac{(8\pi a)^4}{2}c_\text{MS}\notag\\
&=6(D+977.736~695\,a^4).
\end{align*}

At $r_s\ne0$, the discrepancy between Ref.~\cite{Braaten1999} and our result, Eqs.~\eqref{eq:T3}, disappears if $r_s$
is included in the effective field theory. The 2-boson interaction vertex becomes \cite{Braaten2001}
$$\I(-8\pi a-2\pi a^2r_sk^2-2\pi a^2r_sk'^2),$$
where $\vect k_c/2\pm\vect k$ and $\vect k_c/2\pm\vect k'$ are the momenta of the two bosons before and after the interaction,
respectively. $k\ne k'$ if a virtual particle is involved.
The tree diagram contribution to the 3-boson T-matrix, Fig.~5.(a) of Ref.~\cite{Braaten1999}, is modified as
\begin{align*}
&\sum_{i,j=1}^3G^E_{\vect q_j\vect b_i}\big[8\pi a+2\pi a^2r_sp_j^2+2\pi a^2r_s(\vect b_i+\vect q_j/2)^2\big]\\
&\quad\quad\times\big[8\pi a+2\pi a^2r_sh_i^2+2\pi a^2r_s(\vect q_j+\vect b_i/2)^2\big]\\
&=\Big[64\pi^2a^2\sum_{i,j=1}^3G^E_{\vect q_j\vect b_i}+
32\pi^2a^3r_s\sum_{i,j=1}^3(p_j^2+h_i^2)G^E_{\vect q_j\vect b_i}\Big]\\
&\mspace{19mu}+288\pi^2a^3r_s+O(E^1).
\end{align*}
$r_s$ corrections to all other diagrams are $O(E^s)$, $s\ge1/2$.
Comparing the modified EFT results with Eqs.~\eqref{eq:T3}, we get
\begin{equation}\label{eq:g3}
g_3(\lvert a\rvert^{-1})=\frac{6\hbar^2}{m_\text{boson}}(D+12\pi^2a^3r_s+977.736~695\,a^4),
\end{equation}
where the SI units have been restored.

\subsection{Implications for the BEC energy and other properties \label{subsec:BECenergy}}
The BEC energy per particle at zero temperature was computed to order $\rho^2$ by Braaten and Nieto \cite{Braaten1999}.
The result, Eq.~(96) of Ref.~\cite{Braaten1999}, is expressed in terms of $g_3(\kappa)$.

Although the three-boson scattering amplitude receives $r_s$ corrections as shown above,
the BEC energy per particle does not suffer from $r_s$ corrections at order $\rho^2$ in the thermodynamic limit,
if it is expressed in terms of $g_3(\kappa)$ \cite{Braaten2001}.

Using Eq.~\eqref{eq:g3}, we can now express Braaten and Nieto's result for the BEC energy \cite{Braaten1999}
in terms of the scattering hypervolume $D$ defined in Eq.~\eqref{eq:Ddef}.
The outcome is precisely the same as Eq.~\eqref{eq:E0_thermodynamic}, and $C^E$ in Eq.~\eqref{eq:E0_thermodynamic} is found
to be
\begin{align}
C^E&=977.736695/12\pi+8w\big[\ln(16\pi)+0.80\pm0.005\big]\notag\\
&=118.65\pm0.10.\label{eq:CE_Braaten}
\end{align}

The present author did an independent calculation of $C^E$, using finite-range interactions (details to appear elsewhere),
and found
\begin{equation}
C^E=118.498~920~346~444~~~\text{(exactly rounded)},\label{eq:CE}
\end{equation}
which is nearly the same as Eq.~\eqref{eq:CE_Braaten}.

For a dilute Bose gas of hard spheres, using Eqs.~\eqref{eq:DHS}, \eqref{eq:E0_thermodynamic}, \eqref{eq:CE}
and $r_s=2a/3$, we obtain the following result for the constant defined in Eq.~\eqref{eq:E_BEC}:
\begin{equation}\label{eq:E3pHS}
\mathcal{E}_3'=167.319~69\pm0.000~06.
\end{equation}
This completes our calculation of the ground state energy per particle of a dilute Bose gas of hard spheres,
through order $\rho^2$.

For a dilute Bose gas with $a\gg r_e$ ($r_e$ is the range of interparticle forces), Braaten, Hammer, and Mehen
found that $\mathcal{E}_3'$ is near $141$ \cite{Braaten2002} and has a small imiginary part associated with 3-body recombination
\cite{Braaten2002}. In comparison to this system, a hard sphere Bose gas
with the same scattering length $a$, number density $\rho$, and boson mass,
has an energy density that is larger by a relative fraction $\approx26\rho a^3$,
a pressure that is larger by a fraction $\approx52\rho a^3$, a speed of sound that is faster by a fraction
$\approx\frac{3}{2}\times26\rho a^3=39\rho a^3$, and a specific heat that is \emph{smaller} by a fraction
$\approx\frac{9}{2}\times26\rho a^3=117\rho a^3$. Here the specific heats are compared at the same temperature $T\ll\rho^{1/3}aT_c$,
where they are dominated by phonons with wavelengths $\gg1/\sqrt{\rho a}$. ($T_c$ is the critical temperature.)

These differences, \textit{ie} nonuniversal effects, must extend to finite temperatures, including both $T<T_c$ and $T\ge T_c$.
(Their magnitudes will change when $T$ is raised.)

For $T/T_c\sim O(1)$ and $\rho a^3, \rho r_e^3\ll1$, the thermal de Broglie wavelength greatly exceeds $r_e$ and $a$, so the interaction is
well described by the constants $a$, $r_s$, and $D$, etc.

The nonuniversal corrections to $T_c$, as an extension of the calculations
reviewed in Ref.~\cite{Andersen}, are of particular interest.
Is the \emph{leading order} nonuniversal correction determined by $r_s$, or $D$, or both?
The present author speculates it is perhaps by $r_s$ alone, and is of the form $c'\rho a^2r_s T_c^0$, where $T_c^0$ is the
critical temperature of the noninteracting Bose gas, and $c'$ is a universal numerical constant.

\section{Summary\label{sec:conclude}}
We have shown that the three-body force near the scattering threshold
can be predicted in a way very similar to the two-body force, \textit{ie}, by solving
the Schr\"{o}dinger equation at zero energy and matching the solution to the asymptotic
formula for the wave function at large relative distances.
This approach is applicable to the $n$-body force as well.

Although in this paper an explicit calculation of the three-body force is given for hard spheres
only, one can apply the general formulas, Eqs.~\eqref{eq:phi3coordinate}, to many other interaction potentials,
to predict the effective three-body forces. The unknown wave function
has 3 independent variables only, because of the translational and rotational symmetry, and is not
very difficult to study on a present-day computer. 

The author believes that the asymptotics of the three-boson wave function at large relative distances
found in this paper is also applicable to composite bosons with finite-range interactions.
For weakly bound dimers of fermionic atoms, the two-dimer scattering length $a_{d}\approx0.6a_f$ \cite{Petrov},
where $a_f$ is the atomic scattering length, but the parameters $r_s$ and $D$ for these dimers
are still unknown, which must be computed to determine the equation of state
of many such dimers \cite{Petrov,Stringari,TanLevin} more accurately \cite{footnote:vanderWaals}.

We have expanded the ground state energy of three bosons in a large periodic volume to the order $L^{-7}$, using a perturbation procedure
that resembles the derivation of the small-momentum structure of $\phi^{(3)}$.
The result, Eq.~\eqref{eq:Enumerical}, may be combined with a Monte-Carlo simulation (for nonrelativistic particles such as atoms)
or Lattice QCD simulation (for low energy pions) to determine the three-body force.

Equation~\eqref{eq:Enumerical} suggests that, to determine $D$ accurately, $L$ should greatly exceed $17a$ in these simulations, or
the higher order corrections can overwhelm the lower order terms. 

To extract $D$ from $\CN$-body simulations in volumes of modest sizes, one may find it helpful to derive a systematic expansion of $E$ at
large $L$, but fixed $\CN a/L$, corresponding to the \emph{intermediate regime} discussed in Sec.~\ref{subsubsec:EAndx}.

For pions, the Compton wavelength is comparble to $a$ and $r_s$; whether the relativistic corrections
modify the \emph{form} of Eq.~\eqref{eq:Enumerical} is a question of interest.

We have computed the scattering amplitude of three bosons at low energy to $O(E^0)$, using finite-range interactions. Our result, Eqs.~\eqref{eq:T3},
disagrees with an EFT prediction \cite{Braaten1999} at $r_s\ne0$. By including $r_s$ corrections in the EFT, however,
one can eliminate the discrepancy completely.
The resultant relation between $g_3(\kappa)$ and $D$, combined with the many-body energy formula of Ref.~\cite{Braaten1999},
and the three-body force computed in the present paper [Eq.~\eqref{eq:DHS}], solves a longstanding problem of theoretical interest, namely
the complete second-order correction to the ground state energy of a dilute Bose gas of hard spheres. The result shows
small differences between this system and a dilute Bose gas with large scattering length considered by Ref.~\cite{Braaten2002}.

We have studied the energies and momentum distributions of $\CN$ bosons in a large volume,
from which we deduce the energies and condensate fractions of dilute BECs with sizes greatly exceeding the healing length.
The result for the energy, Eq.~\eqref{eq:E0_thermodynamic}, agrees with the effective field theory,
but the condensate depletion, Eq.~\eqref{eq:x_thermodynamic}, disagrees.
It is found that at low density, even the population of the \emph{zero}-momentum state, or the off-diagonal
\emph{long}-range order \cite{YangODLRO}, is affected by short-range physics missed by the EFT.

Although the three-body force is usually a small correction to physical observables in a dilute BEC,
dramatic effects may be obtained near a three-body resonance \cite{Bulgac, Braaten2002}.
Alternatively, one may reduce the two-body force by tuning the scattering length to a zero crossing \cite{ZeroCrossing}.
At $a=0$ the BEC ground state energy density $\approx\hbar^2D\rho^3/6m$.

The author would like to draw the readers' attention to a few mathematical techniques.
The $Z$-functions and $Z$-$\delta$ expansions (Appendices~\ref{sec:Z} and \ref{sec:Z-delta}), an extension of the familiar
$\delta$ function method, facilitate the derivation of the small-momentum and the large-volume expansions.
The method of tail-singularity separation, as described in Appendix~\ref{sec:constants},
enables us to evaluate many lattice sums with virtually arbitrary precision.
The utility of these methods is certainly \emph{not} limited to the present work.

\begin{acknowledgments}
The author thanks N. Barnea, J.-P. Blaizot, A. Bulgac, W. Detmold, Y. Nishida, and M. J. Savage
for useful discussions. This work was supported by Department of Energy grant DE-FG02-00ER41132.
\end{acknowledgments}

\begin{appendix}
\section{The $Z$-functions\label{sec:Z}}
We define generalized functions $Z/k^{2n}$ and $Z_b(k)/k^{2n+1}$
($n=1,2,3,\dots$, and $b>0$ is a constant):
\begin{gather}
\frac{Z}{k^{2n}}=\frac{1}{k^{2n}}\text{~~($k>0$)},~~~\frac{Z_b(k)}{k^{2n+1}}=\frac{1}{k^{2n+1}}\text{~~($k>0$)},\\
\int_\text{all $\vect k$}\dif^3k\frac{Z}{k^{2n}}p^{(s)}(\vect k)=0\text{~~($s\le2n-4$)},\\
\int_{\lvert\vect k\rvert<k_0}\dif^3k\frac{Z}{k^{2n}}p^{(2n-3)}(\vect k)=0\text{~~($k_0>0$ and $n\ge2$)},\\
\int_\text{all $\vect k$}\dif^3k\frac{Z_b(k)}{k^{2n+1}}p^{(s)}(\vect k)=0\text{~~($s\le2n-3$)},\\
\int_{\lvert\vect k\rvert<b}\dif^3k\frac{Z_b(k)}{k^{2n+1}}p^{(2n-2)}(\vect k)=0.
\end{gather}
where $p^{(s)}(\vect k)$ is any homogeneous polynomial of $\vect k$ with degree $s$ ($=0,1,2,\dots$).
$Z/k^2$ can be identified with $1/k^2$.


We can use ordinary functions to approach the $Z$-functions. For instance,
$(k^2-3\eta^2)/(k^2+\eta^2)^3\rightarrow Z/k^4$ when $\eta\rightarrow0^+$.

The $Z$-functions, like the $\delta$ function, are merely Fourier transforms of some ordinary functions ($n=1,2,3,\dots$):
\begin{gather}
\int\frac{\dif^3k}{(2\pi)^3}\frac{Z}{k^{2n}}\e^{\I\vect k\cdot\vect r}=\frac{(-1)^{n+1}r^{2n-3}}{4\pi(2n-2)!},\\
\lim_{\eta\rightarrow0^+}\int\dif^3r\frac{(-1)^{n+1}r^{2n-3}}{4\pi(2n-2)!}\e^{-\I\vect k\cdot\vect r-\eta r}
=\frac{Z}{k^{2n}},\label{eq:rTransform}\\
\int\frac{\dif^3k}{(2\pi)^3}\frac{Z_b(k)}{k^{2n+1}}\e^{\I\vect k\cdot\vect r}=F^{(n)}_b(r),\\
\lim_{\eta\rightarrow0^+}\int\dif^3rF^{(n)}_b(r)\e^{-\I\vect k\cdot\vect r-\eta r}=\frac{Z_b(k)}{k^{2n+1}},\\
F^{(n)}_b(r)\equiv\frac{(-1)^nr^{2n-2}}{2\pi^2(2n-1)!}
\Bigl[\ln(br)+\gamma-\sum\limits_{i=1}^{2n-1}\frac{1}{i}\Bigr],\notag
\end{gather}
where $\gamma=0.5772\cdots$ is Euler's constant.

$Z_b(k)/k^{2n+1}$ has the following properties:
\begin{gather}
\frac{Z_b(k)}{k^{2n+1}}-\frac{Z_{b'}(k)}{k^{2n+1}}=\frac{4\pi\ln(b'/b)}{(2n-1)!}\nabla_k^{2n-2}\delta(\vect k),\label{eq:Z_Shiftb}\\
\frac{Z_b(ck)}{(ck)^{2n+1}}=\frac{1}{c^{2n+1}}\frac{Z_{b/c}(k)}{k^{2n+1}}\text{~~(constant $c>0$)}.
\end{gather}

\section{$Z$-$\delta$ expansions\label{sec:Z-delta}}
\textbf{Example 1.}
Consider the expansion of $(\lvert\vect k\rvert^2+3q^2/4)^{-1}$ at small $q$.
Naively, it is $1/k^2-3q^2/(4k^4)+O(q^4)$. But this series fails around $\vect k=0$.
In particular, $R_0(\vect k,\vect q)=(k^2+3q^2/4)^{-1}-1/k^2$ has a \emph{finite} integral over all $\vect k$:
$\int\frac{\dif^3k}{(2\pi)^3}R_0(\vect k,\vect q)=-\frac{\sqrt{3}}{8\pi}q$,
but the integral of $-3q^2/(4k^4)$ over all $\vect k$ is \emph{infinite}. 

Now subtract $-\frac{\sqrt{3}}{8\pi}q(2\pi)^3\delta(\vect k)$ from $R_0(\vect k,\vect q)$ to obtain
$R_1(\vect k,\vect q)$. Clearly $R_1(\vect k,\vect q)\approx-3q^2/(4k^4)$ at $k>0$ and $q\rightarrow0$,
but $\int R_1(\vect k,\vect q)\dif^3k=0$, so actually
\begin{equation*}
R_1(\vect k,\vect q)\approx-\frac{3q^2}{4}\frac{Z}{k^4},
\end{equation*}
where $Z/k^4$ is defined in Appendix~\ref{sec:Z}. Further subtracting $-3q^2Z/(4k^4)$ from $R_1(\vect k,\vect q)$, we get
a remainder $\approx(\sqrt{3}/64\pi)q^3\nabla_k^2(2\pi)^3\delta(\vect k)$.
Continuing this subtraction procedure to higher orders in $q$, we get Eq.~\eqref{eq:Z-deltaExample1a}. 

The second method to derive the $Z$-$\delta$ expansion of $(k^2+3q^2/4)^{-1}$ is:
first Fourier transform it for fixed $\vect q$,
to obtain $\exp(-\sqrt{3}qr/2)/(4\pi r)$, then expand it in powers of $\vect q$:
$$\frac{1}{4\pi r}-\frac{\sqrt{3}q}{8\pi}+\frac{3q^2}{32\pi}r-\frac{\sqrt{3}q^3}{64\pi}r^2+\frac{3q^4}{512\pi}r^3-\frac{3\sqrt{3}q^5}{5120\pi}r^4
+O(q^6),$$
and finally transform the series back to the $\vect k$-space term by term. With the help of Eq.~\eqref{eq:rTransform},
one gets Eq.~\eqref{eq:Z-deltaExample1a}.

Although the above two methods are equally valid,
the first one is more useful, because the Fourier transforms of most functions of $\vect k$ in this paper
can not be found analytically.

\textbf{Example 2} [needed in deriving Eq.~\eqref{eq:t0-3}].
At small $\vect q$
\begin{align*}
&(\lvert\vect k+\vect q/2\rvert^{-1}+\lvert\vect k-\vect q/2\rvert^{-1})/(k^2+3q^2/4)\\
&\!=\!2Z_{\kappa}(k)/k^3+\big[8\pi-4\pi^2\!/\!\sqrt{3}-8\pi\ln(q/\kappa)\big]\delta(\vect k)+O(q^2),
\end{align*}
where $\kappa>0$ is arbitrary. Higher order corrections may easily be obtained as well
(see the general rules below).

\textbf{Generics.}
Consider a function $F(\vect k,o)$ (where $o$ are a set of variables),
with a finite integral over any finite region of the $\vect k$-space for $o\ne0$, 
and a unique singularity at $\vect k=0$ for $o\rightarrow0$. The $Z$-$\delta$ expansion is of the form
$F(\vect k,o)=F_Z(\vect k,o)+F_\delta(\vect k,o)$, where
$F_Z$ is a series including ordinary and/or $Z$ functions of $\vect k$ and
may be directly inferred from the Taylor expansion of $F(\vect k,o)$ at small $o$, and
\begin{subequations}
\begin{align}
&F_\delta(\vect k,o)=\sum_{mj}c_{mj}(o)\nabla_k^{2m}Q^{(j)}(\nabla_k)(2\pi)^3\delta(\vect k),\\
&c_{mj}(o)=(-1)^{l_j}\big\{\nabla_k^{2m}Q^{(j)}(\nabla_k)\big[k^{2m}Q^{(j)}(\vect k)\big]\big\}^{-1}\notag\\
&\quad\quad\times\int\frac{\dif^3k}{(2\pi)^3}k^{2m}Q^{(j)}(\vect k)\big[F(\vect k,o)-F_Z(\vect k,o)\big].\label{eq:cmj}
\end{align}
\end{subequations}
Here $Q^{(j)}$ are all the independent homogeneous harmonic polynomials, satisfying
$\int Q^{(j)}(\hat{\vect k})Q^{(j')}(\hat{\vect k})\dif^2\hat{\vect k}=0$ for $j\ne j'$.
The degree of $Q^{(j)}$ is $l_j$.

If $F(\vect k,o)$ is rotationally invariant around an axis $\hat{\vect q}$, only $Q^{(l)}_{\hat{\vect q}}$
are needed above, and the coefficient before the integral sign in Eq.~\eqref{eq:cmj} becomes
$\frac{(-1)^l(2l+1)}{(2m)!!(2m+2l+1)!!}$.

If $F(\vect k,o)$ is a completely symmetric and even function of $k_x$, $k_y$, and $k_z$,
only those harmonic polynomials with the $A_1^+$ symmetry are needed:
$Q^{(j)}(\vect k)=1$, $Q^{(g)}(\vect k)$, $Q^{(i)}(\vect k)$, $Q^{(8)}(\vect k)$, \dots, where
\begin{subequations}\label{eq:Qj}
\begin{align}
&Q^{(g)}(\vect v)\equiv v_x^4+v_y^4+v_z^4-3(v_x^2v_y^2+v_y^2v_z^2+v_z^2v_x^2),\\
&\mspace{2mu}Q^{(i)}(\vect v)\equiv v_x^6+v_y^6+v_z^6-({15}/{2})(v_x^4v_y^2+v_y^4v_z^2+v_z^4v_x^2\notag\\
&\mspace{78mu}+v_x^2v_y^4+v_y^2v_z^4+v_z^2v_x^4)+90v_x^2v_y^2v_z^2,\\
&Q^{(8)}(\vect v)\equiv v_x^8+v_y^8+v_z^8-14(v_x^6v_y^2+v_y^6v_z^2+v_z^6v_x^2\notag\\
&+v_x^2v_y^6+v_y^2v_z^6+v_z^2v_x^6)\!+35(v_x^4v_y^4+v_y^4v_z^4+v_z^4v_x^4 ).
\end{align}
\end{subequations}

\section{Lattice sums\label{sec:constants}}
The following methods are used to evaluate lattice sums: \emph{tail-singularity separation}, Poisson summation formula, and
convergence acceleration (based on the large-$\vect n$ asymptotics of the summands).

To evaluate a lattice sum $\sum_\vect nX(\vect n)$ (sum over 3-vectors of integers),
where $X(\vect n)$, as a continuous function, has both singularity in the real-$\vect n$ space and a power-law
tail at large $\vect n$, we sometimes break $X(\vect n)$ in two pieces: $X(\vect n)=X_1(\vect n)+X_2(\vect n)$,
such that $X_1(\vect n)$ has singularity but no power-law tail (ie, decays much more rapidly at large $n$),
while $X_2(\vect n)$ has power-law tail but is sufficiently smooth.
$\sum_\vect nX_1(\vect n)$ is done directly,
while $\sum_{\text{\textbf{all} }\vect n}X_2(\vect n)$ is approximated by $\int\dif^3nX_2(\vect n)$ to a very high precision.
We call this method \emph{tail-singularity separation} (TSS).

For instance, to compute $\alpha_s$ ($s=1,2,3\cdots$), we write
$n^{-2s}=\sum_{i=0}^{s-1}\frac{(\eta n^2)^i}{i!}\frac{\exp(-\eta n^2)}{n^{2s}}+X_2(\vect n)$, where $\eta>0$ is small.
$X_2(\vect n)$ is very smooth, so $\sum_{\text{\textbf{all} }\vect n}X_2(\vect n)$ is approximated by an integral.
Straightforward algebra yields
\begin{align}
\alpha_s&=c_s\pi^{3/2}\eta^{s-3/2}-\frac{\eta^s}{s!}
+\sum_{\vect n\ne0}\sum_{i=0}^{s-1}\frac{(\eta n^2)^i}{i!}\frac{\e^{-\eta n^2}}{n^{2s}}\notag\\
&\quad+O(\e^{-\pi^2/\eta}),\label{eq:alphas_numerical}
\end{align}
where $c_1=-2$, $c_2=2$, $c_3=1/3$, $c_4=1/15$, $c_5=1/84$, $c_6=1/540$, \dots, and the error $\sim O(\e^{-\pi^2/\eta})$ results from the approximation
$\sum_{\text{all }\vect n}X_2(\vect n)\approx\int\dif^3nX_2(\vect n)$ \cite{footnote:TSSerror}.
Equation~\eqref{eq:alphas_numerical} applies to both $s=1$ and $s=2,3,\cdots$.
At $\eta=1/10$, one already gets about 43-digit precision for $\alpha_s$.

The TSS is \emph{not} an arbitrary exponential accelaration method. For example, $\sum_{\vect n\ne0}n^{-3}$
is not sufficiently accelerated by $\sum_{\vect n\ne0}\exp(-\eta n^3)n^{-3}$, because $[1-\exp(-\eta n^3)]n^{-3}$ is \emph{not} smooth enough:
it contains a term $\sim \lvert\vect n\rvert^3$ at small $\vect n$ which is singular. Inspired by 
Takahasi and Mori's quadrature method \cite{footnote:TakahasiMori}, we obtain
a proper TSS formula:
$$1/n^{3}-\pi\eta/2n^2=2/[1+\exp(\pi\sinh\eta n)]n^3+S(\vect n),$$
where $S(\vect n)$ is smooth. We then easily derive (all the lattice sums below, except $\beta_{1A}^{}$, are used in computing
$C_0$ and $C_1$):

\begin{subequations}
\begin{align}
&\alpha_{1.5}^{}=j_1+4\pi\ln\eta+\pi\alpha_1\eta/2+\pi(\pi^2-2)\eta^3/24\notag\\
&+\sum_{\vect n\ne0}2/[1+\exp(\pi\sinh\eta n)]n^3+O(\e^{-\pi^2/\eta}),\\
&j_1\equiv\lim_{R\rightarrow\infty}4\pi\int_{0}^Rx^{-1}\tanh\Big(\frac{\pi}{2}\sinh x\Big)\dif x-4\pi\ln R\notag\\
&\quad\!=16.489~380~548~112~915~838~168~866~783~965~159~811.\notag
\end{align}
Similarly
\begin{align}
&\alpha_{2.5}=\pi\alpha_2\eta/2+j_2\eta^2-\pi(\pi^2-2)\alpha_1\eta^3/24\notag\\
&~-\pi(\pi^4\!-5\pi^2+1)\eta^5/240+\sum_{\vect n\ne0}2/[1+\exp(\pi\sinh\eta n)]n^5\notag\\
&\mspace{202mu}+O(\e^{-\pi^2/\eta}),\\
&j_2\equiv4\pi\int_0^\infty\Big[x^{-3}\tanh\Big(\frac{\pi}{2}\sinh x\Big)-\pi/2x^2\Big]\dif x\notag\\
&=-24.436~776~868~803~521~072~197~485~676~562~043~398.\notag
\end{align}
\end{subequations}

Let
\begin{equation}
\alpha^{(g)}_s\equiv\lim_{\eta\rightarrow0^+}\sum_{\vect n\ne0}\e^{-\eta n}Q^{(g)}(\vect n)/n^{2s},
\end{equation}
and similarly for $\alpha^{(i)}_s$ and $\alpha^{(8)}_s$. Here $Q^{(g)}$, $Q^{(i)}$ and $Q^{(8)}$ are defined
in Eqs.~\eqref{eq:Qj}.

For any \emph{integer} $s\ge1$, we derive a TSS formula
\begin{equation}
\alpha^{(g)}_s=\sum_{\vect n\ne0}\frac{Q^{(g)}(\vect n)}{n^{2s}}\sum_{i=0}^{s-1}\frac{(\eta n^2)^i}{i!}\e^{-\eta n^2}+O(\e^{-\pi^2/\eta}),
\end{equation}
and similarly for $\alpha_s^{(i)}$ and $\alpha_s^{(8)}$.

Using the Poisson summation formula and the TSS method, we get
\begin{subequations}\begin{align}
\alpha^{(g)}_{5.5}&=-(2^6\pi^5/9!!)\!\sum_{\vect n\ne0}\!Q^{(g)}(\vect n)\widetilde{l}(\eta n)\!+\!O(\e^{-\pi^2/\eta}),\\
\alpha^{(i)}_{7.5}&=(2^8\pi^7/13!!)\sum_{\vect n\ne0}Q^{(i)}(\vect n)\widetilde{l}(\eta n)+O(\e^{-\pi^2/\eta}),\\
\alpha^{(8)}_{9.5}&=-(2^{10}\pi^9/17!!)\sum_{\vect n\ne0}Q^{(8)}(\vect n)\widetilde{l}(\eta n)\!+\!O(\e^{-\pi^2/\eta}),
\end{align}\end{subequations}
where $\widetilde{l}(x)\equiv\ln\!\big[\!\tanh\!\big(\frac{\pi}{2}\sinh x\big)\!\big]$.

Using the Poisson summation formula, we get
\begin{subequations}\label{eq:rho_evaluate}
\begin{align}
W_A(\vect n)&=-\sqrt{3}\,\pi^2n+\pi\sum_{\vect m\ne0}\e^{-\sqrt{3}\,\pi m n-\I\pi\vect m\cdot\vect n}/m,\label{eq:WA_evaluate}\\
W_B(\vect n)&=(2\pi^2/\sqrt{3}\,n)\sum_{\text{all }\vect m}\e^{-\sqrt{3}\,\pi m n-\I\pi\vect m\cdot\vect n}.
\end{align}
We derive TSS formulas
\begin{align}
&\theta_{A1}(\vect n)=-\eta(1-\e^{-\eta n^2})/n^2+2\pi\int_0^\infty\dif m\int_{-1}^1\dif c\notag\\
&\quad\times(1-\e^{-\eta S_{nmc}})(1-\e^{-\eta m^2})/S_{nmc}+\sum_{\vect m\ne0}S_{\vect n\vect m}^{-1}m^{-2}\notag\\
&\times\!(\e^{-\eta S_{\vect n\vect m}}+\e^{-\eta m^2}\!\!-\!\e^{-\eta S_{\vect n\vect m}-\eta m^2})+O(\e^{-\pi^2/2\eta}),\\
&\theta_{B1}(\vect n)=\eta\big[(1+\eta n^2)\e^{-\eta n^2}-1\big]/n^4+2\pi\int_0^\infty\dif m\int_{-1}^1\dif c\notag\\
&\quad\times\big[1-(1+\eta S_{nmc})\e^{-\eta S_{nmc}}\big](1-\e^{-\eta m^2})/S^{2}_{nmc}\notag\\
&+\sum_{\vect m\ne0}S_{\vect n\vect m}^{-2}m^{-2}\big[(1\!+\!\eta S_{\vect n\vect m})
\e^{-\eta S_{\vect n\vect m}}(1\!-\!\e^{-\eta m^2})+\e^{-\eta m^2}\big]\notag\\
&\quad\quad\quad\quad\quad\quad\quad\quad\quad\quad\quad\quad\quad\quad+O(\e^{-\pi^2/2\eta}),
\end{align}
\end{subequations}
where $S_{\vect n\vect m}\equiv n^2+\vect n\cdot\vect m+m^2$, and $S_{nmc}\equiv n^2+nmc+m^2$. With Eqs.~\eqref{eq:rho_evaluate},
$\rho_{A1}^{}(\vect n)$ and $\rho_{B1}^{}(\vect n)$ [Eqs.~\eqref{eq:rhoA1_def} and \eqref{eq:rhoB1_def}] are evaluated
very accurately for any \emph{finite} 3-vector of integers $\vect n\ne0$.

In addition, at \emph{large} $\vect n$ we derive asymptotic formulas
\begin{subequations}
\begin{align}
&\rho_{A1}^{}(\vect n)=\rho_{A1}^{(11)}(\vect n)+O(n^{-12}),\label{eq:rhoA1_asymp}\\
&\rho_{A1}^{(11)}(\vect n)\!\equiv\pi^2w/n\!+2\alpha_1/n^2\!+\!{4}/3n^4\!+\!{4}\alpha^{(g)}_1Q^{(g)}(\vect n)/15n^{10}\notag\\
&\quad\quad\quad\quad+{4}\alpha^{(i)}_1Q^{(i)}(\vect n)/231n^{14}+{4}\alpha^{(8)}_1Q^{(8)}(\vect n)/195n^{18},\notag\\
&\rho_{B1}^{}(\vect n)=\rho_{B1}^{(13)}(\vect n)+O(n^{-14}),\label{eq:rhoB1_asymp}\\
&\rho_{B1}^{(13)}(\vect n)\!\equiv2\sqrt{3}\,\pi^2\!/n^3\!\!+\!2\alpha_1/n^4\!\!+2/n^6\!\!+\!4\alpha^{(g)}_1Q^{(g)}(\vect n)/3n^{12}\notag\\
&\quad\quad\quad\quad+4\alpha^{(i)}_1Q^{(i)}(\vect n)/33n^{16}+12\alpha^{(8)}_1Q^{(8)}(\vect n)/65n^{20}.\notag
\end{align}
\end{subequations}

For $\alpha_{1A1}^{}$, $\alpha_{2A1}^{}$, and $\alpha_{1B1}^{}$, we thus have
\begin{subequations}
\begin{align}
\alpha_{sA1}^{}&=\pi^2w\alpha_{0.5+s}^{}+2\alpha_1^{}\alpha_{1+s}^{}+4\alpha_{2+s}^{}/3+4\alpha^{(g)}_1\!\alpha^{(g)}_{5+s}/15\notag\\
&\quad+4\alpha^{(i)}_1\alpha^{(i)}_{7+s}/231+4\alpha^{(8)}_1\alpha^{(8)}_{9+s}/195\notag\\
&\quad+\sum_{\vect n\ne0}n^{-2s}\big[\rho_{A1}^{}(\vect n)-\rho_{A1}^{(11)}(\vect n)\big],\\
\alpha_{1B1}^{}&=2\sqrt{3}\,\pi^2\alpha_{2.5}^{}+2\alpha_1^{}\alpha_3^{}+2\alpha_4^{}+4\alpha^{(g)}_1\alpha^{(g)}_7/3\notag\\
&\quad+4\alpha^{(i)}_1\alpha^{(i)}_9/33+12\alpha^{(8)}_1\alpha^{(8)}_{11}/65\notag\\
&\quad+\sum_{\vect n\ne0}n^{-2}\big[\rho_{B1}^{}(\vect n)-\rho_{B1}^{(13)}(\vect n)\big],
\end{align}
\end{subequations}
and the sums over $\vect n$ are truncated at $\lvert n_{x}\rvert, \lvert n_{y}\rvert, \lvert n_{z}\rvert\le15$
to yield results for $\alpha_{1A1}^{}$, $\alpha_{2A1}^{}$, and $\alpha_{1B1}^{}$ with more than 18-digit precision. 

From Eq.~\eqref{eq:rhotAA1_def} we deduce
\begin{align}
&\widetilde{\alpha}_{1AA1}^{}=-\alpha_{1AA1}^{}-\pi^3w(7\pi/\sqrt{3}-8)\alpha_1^{}+8\pi^3w\alpha_1^{}\ln(2\pi)\notag\\
&-\!\alpha_1^{}\alpha_{1A1}^{}\!+3\alpha_{1A2}^{}\!+3\alpha_{1B1}^{}\!-\!(\alpha_1^2+\alpha_2^{})\alpha_2^{}+6\alpha_1^{}\alpha_3^{}\!-9\alpha_4^{},\\
&\alpha_{1AA1}^{}\equiv\lim_{N\rightarrow\infty}\sum_{\vect n\ne0;~n<N}n^{-2}\rho_{AA1}^{}(\vect n)-8\pi^3w\alpha_1\ln N\notag\\
&\quad\quad\quad\quad\quad+4\pi^4wN\big(8\ln N-16+7\pi/\sqrt{3}\,\big).
\end{align}
It can be shown that $\alpha_{1A2}^{}=\alpha_{2A1}^{}$, and
\begin{equation}
\alpha_{1AA1}^{}=\!\lim_{N\rightarrow\infty}\!\sum_{\vect n\ne0;n<N}\!\!\!\!\!\rho_{A1}^2(\vect n)-4\pi^5w^2N-16\pi^3w\alpha_1^{}\ln N.
\end{equation}
At large $\vect n$ we derive from Eq.~\eqref{eq:rhoA1_asymp}
\begin{equation}
\rho_{A1}^2(\vect n)=K^{(12)}(\vect n)+O(n^{-13}),
\end{equation}
\begin{widetext}
\begin{multline}
K^{(12)}(\vect n)\equiv\pi^4w^2/n^2+4\pi^2w\alpha_1/n^3+4\alpha_1^2/n^4+8\pi^2w/3n^5+16\alpha_1/3n^6
+{8}\pi^2w\alpha^{(g)}_1Q^{(g)}(\vect n)/15n^{11}\\
+\big[16/9n^8+16\alpha_1\alpha^{(g)}_1Q^{(g)}(\vect n)/15n^{12}\big]+8\pi^2w\alpha^{(i)}_1Q^{(i)}(\vect n)/231n^{15}
+\big[32\alpha^{(g)}_1Q^{(g)}(\vect n)/45n^{14}+16\alpha_1\alpha^{(i)}_1Q^{(i)}(\vect n)/231n^{16}\big]\\
+8\pi^2w\alpha^{(8)}_1Q^{(8)}(\vect n)/195n^{19}
+\big[{64}\alpha^{(g)2}_1/4725n^{12}+{64}\alpha^{(g)2}_1Q^{(g)}(\vect n)/3575n^{16}
+\big({128}\alpha^{(g)2}_1/10395+{32}\alpha^{(i)}_1/693\big)Q^{(i)}(\vect n)/n^{18}\\
+\big({16}\alpha^{(g)2}_1/585+{16}\alpha_1\alpha^{(8)}_1/195\big)Q^{(8)}(\vect n)/n^{20}\big].
\end{multline}
So
\begin{multline}
\alpha_{1AA1}^{}=\pi^4w^2\alpha_1+4\pi^2w\alpha_1^{}\alpha_{1.5}^{}+4\alpha_1^2\alpha_2^{}+{8}\pi^2w\alpha_{2.5}^{}/3
+16\alpha_1\alpha_3^{}/3+{8}\pi^2w\alpha^{(g)}_1\alpha^{(g)}_{5.5}/15
+16\alpha_4^{}/9+{16}\alpha_1^{}\alpha^{(g)}_1\alpha^{(g)}_6/15\\
+{8}\pi^2w\alpha^{(i)}_1\alpha^{(i)}_{7.5}/231
+{32}\alpha^{(g)}_1\alpha^{(g)}_7/45+{16}\alpha_1\alpha^{(i)}_1\alpha^{(i)}_8/231+{8}\pi^2w\alpha^{(8)}_1\alpha^{(8)}_{9.5}/195
+{64}\alpha^{(g)2}_1\alpha_6^{}/4725+{64}\alpha^{(g)2}_1\alpha^{(g)}_8/3575\\
+\big({128}\alpha^{(g)2}_1/10395+32\alpha^{(i)}_1/693\big)\alpha^{(i)}_9
+\big({16}\alpha^{(g)2}_1/585+{16}\alpha_1\alpha^{(8)}_1/195\big)\alpha^{(8)}_{10}
+\sum_{\vect n\ne0}\big[\rho_{A1}^2(\vect n)-K^{(12)}(\vect n)\big],
\end{multline}
\end{widetext}
where the sum over $\vect n$ converges very rapidly;
truncating it at $\lvert n_{x}\rvert, \lvert n_{y}\rvert, \lvert n_{z}\rvert\le15$ we obtain the result for $\alpha_{1AA1}^{}$ with 18-digit precision.

The results for the above lattice sums are listed below. They determine $C_0$ and $C_1$.

1-fold lattice sums (rounded to fit the 2-column format):
\begin{subequations}
\begin{align}
\alpha_1&=-8.913~632~917~585~151~272~687~120~136,\\
\alpha_{1.5}&=3.821~923~503~940~635~799~730~123~034,\\
\alpha_2&=16.532~315~959~761~669~643~892~704~593,\\
\alpha_{2.5}&=10.377~524~830~847~083~864~728~948~355,
\end{align}
\begin{align}
\alpha_3&=8.401~923~974~827~539~993~146~138~987,\\
\alpha_4&=6.945~807~927~226~369~624~170~778~023,\\
\alpha_6&=6.202~149~045~047~518~551~930~416~392,\\
\alpha^{(g)}_1&=1.127~757~148~686~792~075~014~583~731,\\
\alpha^{(g)}_{5.5}&=5.672~605~625~422~259~129~524~572~370,\\
\alpha^{(g)}_6&=5.772~772~158~341~296~181~095~291~055,\\
\alpha^{(g)}_7&=5.890~866~300~404~517~457~312~864~285,\\
\alpha^{(g)}_8&=5.947~443~615~615~246~912~541~646~623,
\end{align}
\begin{align}
\alpha^{(i)}_1&=-2.434~385~049~385~522~287~574~679~853,\\
\alpha^{(i)}_{7.5}&=5.242~702~841~443~828~214~862~689~624,\\
\alpha^{(i)}_8&=5.451~072~910~162~494~576~739~294~958,\\
\alpha^{(i)}_9&=5.715~464~651~152~249~495~454~861~607,\\
\alpha^{(8)}_1&=16.571~410~717~493~131~178~469~668~510,\\
\alpha^{(8)}_{9.5}&=6.156~579~477~795~506~959~879~789~010,\\
\alpha^{(8)}_{10}&=6.109~685~927~989~065~132~252~619~095,\\
\alpha^{(8)}_{11}&=6.054~088~533~916~337~242~283~438~742.
\end{align}
\end{subequations}

2-fold lattice sums:
\begin{subequations}
\begin{align}
\alpha_{1A1}^{}&=-190.172~897~984~865~754~8,\\
\alpha_{1A2}^{}&=\alpha_{2A1}^{}=111.807~832~628~721~133~609,\\
\alpha_{1B1}^{}&=221.523~005~657~695~107~22,
\end{align}
\end{subequations}

3-fold lattice sums:
\begin{subequations}
\begin{align}
\alpha_{1AA1}^{}&=-2996.889~395~378~764~86,\\
\widetilde{\alpha}_{1AA1}^{}&=-6591.229~842~103~439~89.
\end{align}
\end{subequations}

Finally, from Eq.~\eqref{eq:WA_evaluate} we get
\begin{align}
\beta_{1A}&=-\sqrt{3}\,\pi^2\alpha_{0.5}^{}+\pi\sum_{\vect n,\vect m\ne0}\e^{-\sqrt{3}\,\pi m n-\I\pi\vect m\cdot\vect n}/n^2m\notag\\
&=48.614~754~175~227~821~038~934~419~912.\label{eq:beta1A_numerical}
\end{align}
(One can show that $\alpha_{0.5}^{}=\alpha_1^{}/\pi$.)
$\beta_{1A}^{}$ is needed in Eq.~\eqref{eq:ANNNresult}.

\end{appendix}
\bibliography{ThreeBosonRefs}

\begin{thebibliography}{36}
\bibitem[{not()}]{note:Absa}
\bibinfo{note}{The $\lvert a\rvert$ in $\ln(q_i\lvert a\rvert)$ ensures that
  Eq.~\eqref{eq:Ddef} is also valid for $a<0$.}
\bibitem[{foo({\natexlab{c}})}]{footnote:specialfunctions}
\bibinfo{note}{These functions can be obtained from the wave function of the
  two-body $l$-wave scattering state at energy $E\rightarrow0$:
  $\phi^{(l,E)}_{\hat{\vect n}\vect k}=\phi^{(l)}_{\hat{\vect n}\vect k}
  +Ef^{(l)}_{\hat{\vect n}\vect k}+E^2g^{(l)}_{\hat{\vect n}\vect k}+O(E^3)$,
  whose amplitude is defined by
  $\int\frac{\dif^3k}{(2\pi)^3}\phi^{(l,E)}_{\hat{\vect n}\vect k}\e^{\I\vect
  k\cdot\vect r}= (-1)^{l+1}a_lQ^{(l)}_{\hat{\vect
  n}}(\nabla_r)\big[\sin(\sqrt{E}r+\delta_l)/r\sin\delta_l\big]$ for $r$
  greater than the range of the interaction. Expanding the latter expression in
  powers of $E$, and using Eq.~\eqref{eq:phaseshift}, one can derive
  Eqs.~\eqref{eq:2bodyspecial}. Here $Q^{(l)}_{\hat{\vect n}}$ is defined by
  Eq.~\eqref{eq:Qdirection}.}
\bibitem[{foo({\natexlab{d}})}]{footnote:deltak}
\bibinfo{note}{Note that $Q^{(l)}_{\hat{\vect n}}(\nabla_k)(2\pi)^3\delta(\vect
  k)$ scales with $k$ like $k^{-3-l}$.}
\bibitem[{foo({\natexlab{e}})}]{footnote:a_rs_real}
\bibinfo{note}{It can be shown that $a$ and $r_s$ are real.}
\bibitem[{foo({\natexlab{f}})}]{footnote:phi3}
\bibinfo{note}{The vector sum of the three subscripts of $\phi^{(3)}$ is fixed
  as zero.}
\bibitem[{foo({\natexlab{g}})}]{footnote:dk}
\bibinfo{note}{Using the expansion formulas for $\phi_\vect k$ and
  $\phi^{(3)}_{-\vect k,\vect k/2+\vect k',\vect k/2-\vect k'}$ at small $\vect
  k$, and noting that $U^\phi_{\vect k,-\vect k,0}$ and $\int_{\vect
  k'}U_{\vect k\vect k'}d_{\vect k'}$ are smooth functions of $\vect k$, one
  can readily verify that Eqs.~\eqref{eq:d1} and \eqref{eq:d2} are compatible.}
\bibitem[{foo({\natexlab{i}})}]{footnote:alpha_difference}
\bibinfo{note}{Our numerical answers for $\alpha_1^{}$ and $\alpha_3^{}$ very
  slightly differ from those provided in Ref.~\cite{Savage} by 25 parts per
  trillion and 47.0 parts per trillion, respectively.}
\bibitem[{foo({\natexlab{j}})}]{footnote:u0independence}
\bibinfo{note}{By studying small perturbations of the hard sphere potential at
  $r>a$, one can at least show rigorously that $\Re u_0$ is not a function of
  any finite collection of s-wave parameters $a, r_s, r_s', \dots$ and $l$-wave
  parameters $a_l$'s.}
\bibitem[{foo({\natexlab{l}})}]{footnote:Psik1k2k3}
\bibinfo{note}{It can be shown that the leading correction to
  Eq.~\eqref{eq:Psik1k2k3} is $\Psi_{\vect k_1\vect k_2\vect
  k_3}-\phi^{(3)}_{\vect k_1\vect k_2\vect k_3}\approx -\I
  a(h_1+h_2+h_3)\phi^{(3)}_{\vect k_1\vect k_2\vect k_3}$, but the next
  correction $\sim E^1$ is no longer proportional to $\phi^{(3)}_{\vect
  k_1\vect k_2\vect k_3}$.}
\bibitem[{foo({\natexlab{m}})}]{footnote:vanderWaals}
\bibinfo{note}{Caution should be exercised in dealing with ultracold bosonic
  atoms and molecules, since the van der Waals potential is not strictly
  finite-ranged, but has a tail $-C_6/r^6$ at large $r$, which affects
  $\phi^{(3)}(\vect r_1\vect r_2\vect r_3)$ at the order $R^{-4}$, the
  \emph{same} order as $D$. Effects of this tail on the three-body scattering
  amplitude and the BEC energy may be comparable to $D$, and are currently
  under investigation.}
\bibitem[{foo({\natexlab{n}})}]{footnote:TSSerror}
\bibinfo{note}{The error can be deduced from the Poisson summation formula.}
\end{thebibliography}
\end{document}